\documentclass[journal]{IEEEtran}
\usepackage{cite}

\ifCLASSINFOpdf
  \usepackage[pdftex]{graphicx}
  \graphicspath{{../pdf/}{../jpeg/}{../figures/}}
\else
\fi
\ifCLASSOPTIONcompsoc
\usepackage[caption=false, font=normalsize, labelfont=sf, textfont=sf]{subfig}
\else
\usepackage[caption=false, font=footnotesize]{subfig}
\usepackage{booktabs}
\usepackage{multirow}
\usepackage{color}

\usepackage{amsmath}
\usepackage{amsthm}
\usepackage{algorithm}
\usepackage{algorithmic}

\usepackage{array}

\ifCLASSOPTIONcompsoc
 \usepackage[caption=false,font=normalsize,labelfont=sf,textfont=sf]{subfig}
\else
 \usepackage[caption=false,font=footnotesize]{subfig}
\fi
\usepackage{fixltx2e}
\usepackage{dblfloatfix}

\ifCLASSOPTIONcaptionsoff
 \usepackage[nomarkers]{endfloat}
\let\MYoriglatexcaption\caption
\renewcommand{\caption}[2][\relax]{\MYoriglatexcaption[#2]{#2}}
\fi
\usepackage{url}

\begin{document}
%
\title{Efficient Scheduling of Discrete Industrial Processes through Continuous Modeling}
%
%
%

\author{Ruike Lyu,~\IEEEmembership{Student Member,~IEEE}, Xiangbo Su, Ershun Du,~\IEEEmembership{Member,~IEEE}, Hongye Guo$^*$,~\IEEEmembership{Member,~IEEE}, Qixin Chen,~\IEEEmembership{Senior Member,~IEEE}, and Chongqing Kang,~\IEEEmembership{Fellow,~IEEE}

\thanks{This work was supported by the National Key R\&D Program of China (No. 2023YFB2407300).

R. Lyu, X. Su, H. Guo, Q. Chen, and C. Kang are with the State Key Laboratory of Power System Operation and Control, Department of Electrical Engineering, Tsinghua University, Beijing 100084, China. E. Du is with the Lab of Low Carbon Energy, Tsinghua University,
Beijing 100084, China. (Corresponding author: Hongye Guo. Email: hyguo@tsinghua.edu.cn).}}

\maketitle

\begin{abstract}
  The resource-task network (RTN) model has been widely applied to represent the technical constraints of complex industrial processes (IPs) such as steel-making, providing the basis for industrial demand response. However, the legacy RTN model contains numerous binary variables and applies different formulations for non-flexible and flexible processes, restricting its computational efficiency and applicability. To systematically improve the computational performance of IP models, we propose continuous RTN model (cRTN), a novel modeling approach that uses continuous variables to represent production tasks and progresses, which are then integrated into unified as well as computationally favorable formulations for the technical constraints in discrete IPs, including resource balance, task execution, waiting time limits, and production targets. Compared to the legacy models, cRTN features fewer binary variables, shorter solving time, and better scalability while maintaining the same accuracy. Numerical tests based on a steel plant demonstrate that cRTN is in typical cases 10 times faster than legacy models and remains tractable with increasing batch sizes, which in legacy models leads to larger problem scales and infeasible solving time. cRTN also achieves a reduction in energy costs by resolving the issue of rounding errors reported in legacy models.
\end{abstract}


\ifCLASSOPTIONpeerreview
  \begin{center} \bfseries EDICS Category: 3-BBND \end{center}
\fi
%
\IEEEpeerreviewmaketitle
\section*{Nomenclature}
\addcontentsline{toc}{section}{Nomenclature}

\subsection*{The Legacy RTN Model}
\begin{IEEEdescription}[\IEEEusemathlabelsep\IEEEsetlabelwidth{\(P^{L}_{i}\), \(P^{H}_{i}\)}]
\item[\(\delta\)] Length of time slots.
\item[\(E^{\rm req}_i\)] Required energy consumption to complete task \(i\).
\item[\(i\)] Index of tasks.
\item[\(i+\)] Index of the subsequent task of task \(i\).
\item[\(N_{i, t}\)] Binary variable modeling the start of task \(i\) at time slot \(t\).
\item[\(P^{L}_{i}\), \(P^{H}_{i}\)] Minimum and maximum processing powers of task \(i\), respectively.
\item[\(P_{i, t}\)] Processing power of task \(i\) at time slot \(t\).
\item[\(\Pi_{E L, t}\)] Energy consumption of all production processes at time \(t\).
\item[\(Pr_{t}\)] Electricity price at time slot \(t\).
\item[\(r\)] Index of the resource.
\item[\(r^{\rm d}\)] Index of the product located at the transfer destination.
\item[\(r^{\rm f}\)] Index of the product located at the final stage.
\item[\(r^{\rm s}\)] Index of the product located at the transfer start point.
\item[\(R_{r, t}\)] Value of resource \(r\) at time $t$.
\item[\(S_{i, t}\)] Processing status of task \(i\) at time slot \(t\).
\item[\(t\)] Index of discrete time slots.
\item[\(\tau^L_i\), \(\tau^H_i\)] Minimum and maximum processing times of task \(i\), respectively.
\item[\(W_{r^{\rm d}}\)] Maximum waiting time of resource \(r^{\rm d}\).
\item[\(w_{r^{\rm d}}\)] Transfer time of resource \(r^{\rm d}\).
\item[\(\mu_{E L, i, \theta}\)] Energy consumption of task \(i\), \(\theta\) time slots after the task starts.
\item[\(\mu_{r, i, \theta}\)] Amount of resource \(r\) consumed/generated by task \(i\), \(\theta\) time slots after the start of task \(i\).
\end{IEEEdescription}

\subsection*{Continuous RTN formulation}
\begin{IEEEdescription}[\IEEEusemathlabelsep\IEEEsetlabelwidth{\(P^{L}_{i}\), \(P^{H}_{i}\)}]
\item[\(D_{i, k, t}\)] Time duration that task \(i\) operates in state \(k\) within time slot \(t\).
\item[\(G\)] Task-resource association matrix.
\item[\(g_{r, i, k}\)] Changed amount (generated or consumed) of resource \(r\) when task \(i\) operates in state \(k\) per unit time.
\item[\(i^{\rm end}\)] Final product index.
\item[\(k\)] The state index of the task.
\item[\(R^{\rm tg}_{i^{\rm end}}\)] Target amount of the final product.
\item[\(R_{r^{i+}, t}\)] Progress of the task that generates $r^{i+}$
\item[\(r^{i+}\)] Index of the resource produced by task \(i\).
\item[\(r^{i-}\)] Index of the resources consumed by task \(i\).
\item[\(u_{i, t}\)] Binary variables to characterize the processing state of the task, where \(u_{i, t}=1\) represents that the current process is ongoing.
\item[\(w_i\), \(W_i\)] Transfer time and maximum waiting time of task \(i\), respectively.
\item[\(P_{i, k}\)] Power consumption of task \(i\) operating in state \(k\).
\end{IEEEdescription}

\section{Introduction}
%
%
%
%
\IEEEPARstart{T}{o} address the threats of energy crisis and climate change, countries around the world are setting zero-carbon transition goals for their energy systems~\cite{zhuo_cost_2022}. In this context, coal or natural gas power plants are being replaced by renewable energies such as wind and solar power~\cite{chen_pathway_2021}. As an inevitable aspect of the process, decarbonizing power systems will face unprecedented challenges of insufficient flexibility. In the context of power system operation, flexibility refers to the ability of generation or consumer side resources to change their power exchange with the grid according to the needs of the system, which is crucial for maintaining real-time supply-demand balance~\cite{akrami_power_2019}. However, the availability of renewable energy like wind and solar is dependent on weather conditions and their output cannot be flexibly adjusted as conventional generators~\cite{zhang_data-driven_2023}. Therefore, future power systems must explore and utilize demand-side flexibility to secure power balance~\cite{mohandes_incentive_2021}.

\begin{figure}[!t]
  \centering
  \includegraphics[width=2.0in]{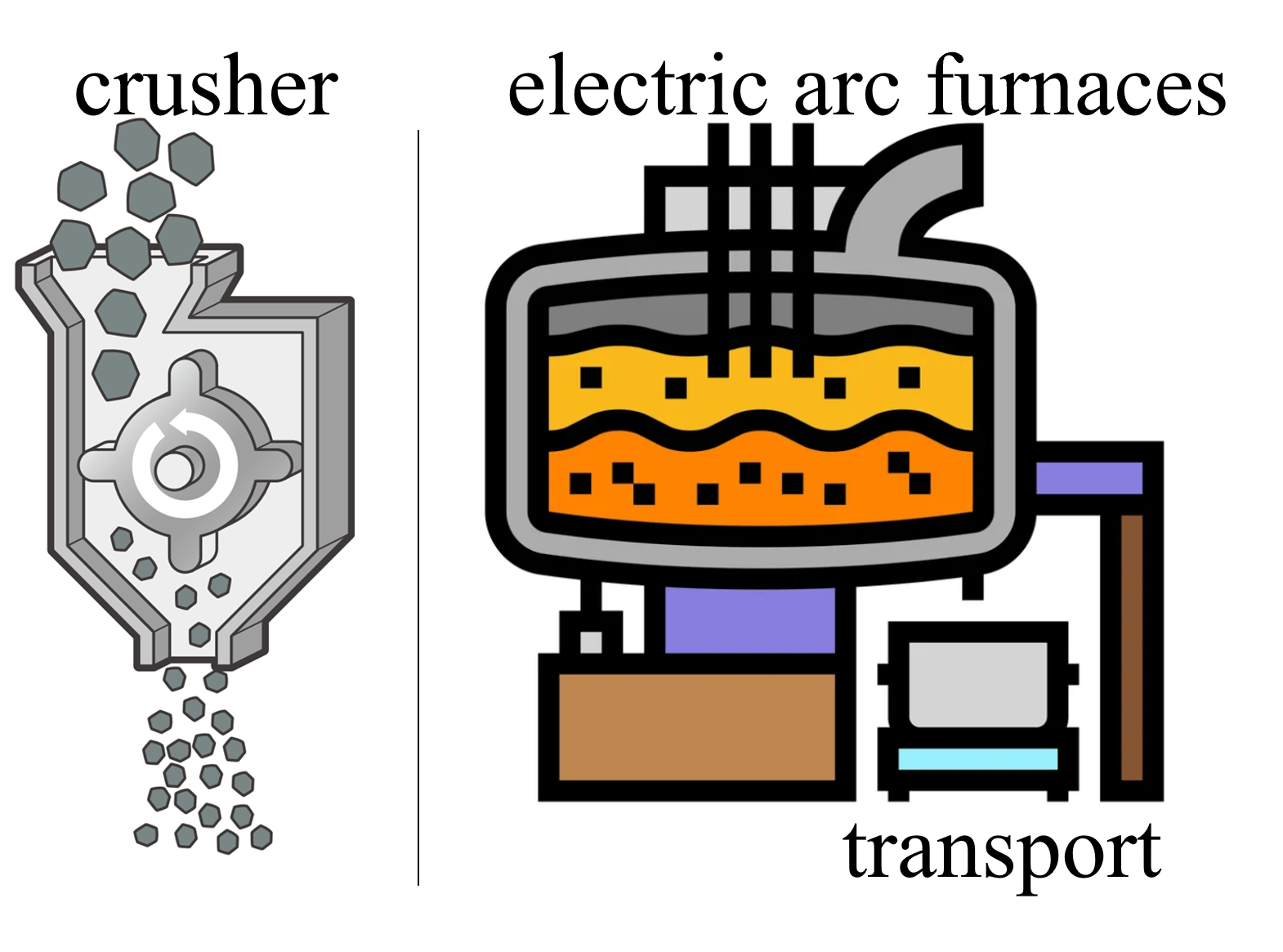}
  \caption{Typical continuous (left) and discrete (right) production processes.}
  \label{fig_discrete_ipp}
\end{figure}

\begin{figure}[!t]
  \centering
  \subfloat[]{
    \includegraphics[width=1.75in]{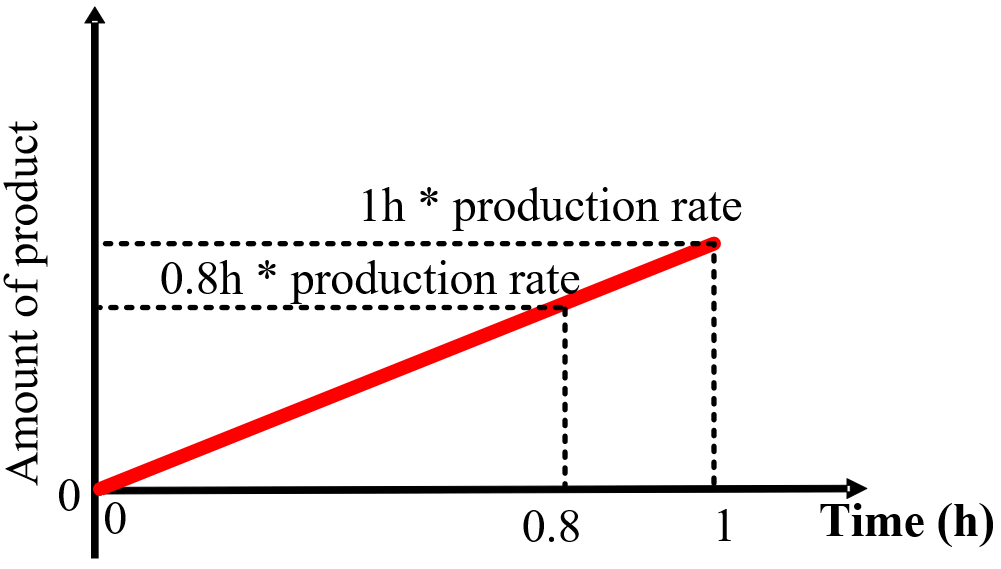}}
  \subfloat[]{
    \includegraphics[width=1.75in]{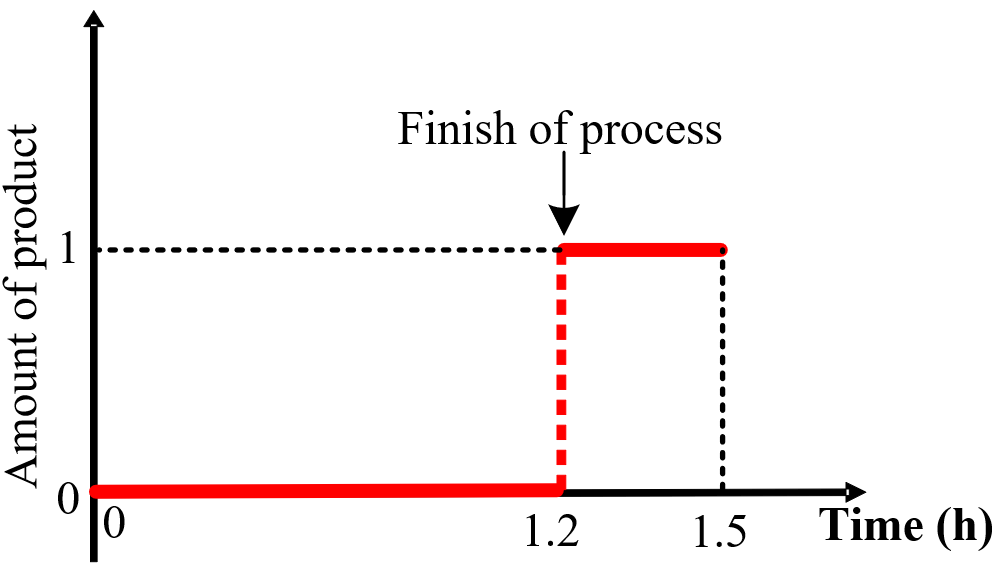}}
  \caption{The quantity of products in a continuous production process (a) is a continuous function of time, while in a discrete production process (b), products are only available for subsequent stages after the entire processing procedure is completed.}
  \label{fig_discrete_ipp2}
\end{figure}

Industrial loads are an ideal demand-side resource for providing flexibility to the power grid through demand response, that is, changing energy consumption or reducing the load in response to electricity price signals~\cite{zhang_cost-effective_2017} or instructions from the grid~\cite{zhang_demand_2018}. There are several advantages of industrial demand response over residential counterparts. First, industrial consumers have substantial response capacities, typically in the megawatt range (e.g., 90MW for steel production lines~\cite{castro_resourcetask_2013}), far exceeding those of residential consumers in the kilowatt range~\cite{shafie-khah_comprehensive_2019}. Larger capacity significantly reduces the transaction and communication costs per megawatt of response capacity which would otherwise require the aggregation of hundreds of residential consumers. Second, typical heavy industrial consumers demonstrate high price sensitivity since electricity costs can comprise 40\% of their operating expenses~\cite{golmohamadi_demand-side_2022}. Third, many industrial consumers already have installed sophisticated automated control systems for their production processes, enabling precise and reliable operations for demand response~\cite{samad_smart_2012}. The significance of industrial demand response is evident across different regions: in Germany, industrial DR potential reaches 10GW (13\% of national peak load)~\cite{german_energy_agency_2018}, while in the United States, industrial consumers contributed 45.6\% of the total 12GW peak reduction despite representing only 19\% of electricity consumption~\cite{ferc_assessment_2023}.

For typical industrial consumers, the production processes accounts for the majority of the overall energy usage and serves as the most crucial source of flexibility for demand response~\cite{golmohamadi_demand-side_2022}. Specifically, based on the technical characteristics, typical IPs can be classified into two categories: continuous IPs and discrete IPs (Figs.~\ref{fig_discrete_ipp}-~\ref{fig_discrete_ipp2}). In continuous IPs, such as the steel powder produced by a crusher, the amount of products can be represented as a continuous function of processing duration~\cite{lu_data-driven_2021}. In contrast, in discrete IPs, products are not continuously produced. For example, in a steel plant, an electric arc furnace (EAF) takes 80 minutes to completely melt a batch of steel, and no molten steel can be fed to the next process before then~\cite{castro_resourcetask_2013}. Notably, discrete IPs often exhibit larger power capacity and greater flexibility compared to continuous IPs. One reason can be that discrete IPs typically involve thermal and chemical processes that require more energy than the purely mechanical processes common in continuous IPs. For instance, in secondary steel-making and aluminum electrolysis, the thermal inertia of the processes allows for flexible power adjustment within certain time windows without significantly impacting product quality~\cite{zhang_bidding_2015}. Therefore, this paper mainly focuses on discrete IPs and studies the mathematical modeling of their technical constraints, which is the basis for minimizing the negative impact on industrial consumers in demand response~\cite{ding_demand_2014}.

\begin{table*}[htbp]
  \caption{Comparison of Continuous IPs and Discrete IPs}
  \label{tab:ipp_comparison}
  \centering
  \begin{tabular}{p{2.5cm}|p{6cm}|p{6cm}}
  \toprule
  \textbf{Characteristic} & \textbf{Continuous IPs} & \textbf{Discrete IPs} \\
  \midrule
  \textbf{Production Process} & 
  • Continuous flow without minimum size limit\newline
  • Can process materials in arbitrary quantities (e.g., 1kg/min of steel powder)\newline
  • Naturally modeled with continuous variables & 
  • Batch-based processing with minimum batch size\newline
  • Must complete the entire batch (e.g., 6 tons of steel)\newline
  • Requires binary variables for state description \\
  \midrule
  \textbf{Operating States} & 
  • Discrete states (e.g., on/off)\newline
  • Flexible adjustment of operating time within states\newline
  • Flexible state transitions & 
  • Discrete states (e.g., on/off)\newline
  • Operating time coupled with production progress\newline
  • State transition constrained by material flow \\
  \midrule
  \textbf{Stage Coupling} & 
  • Modeled through material transformation matrices\newline
  • Linear storage constraints\newline
  • Continuous material flow between stages & 
  • Strict sequential processing requirements\newline
  • Batch completion needed before being processed by following stagess\newline
  • Time-sensitive transfer constraints between stages \\
  \bottomrule
  \end{tabular}
  \end{table*}

Proposed by Pantelides et al.\cite{pantelides1994unified}, the resource-task network (RTN) models manufacturing devices and products as resources and models processes of production and transportation as tasks. In a discrete time horizon, RTN can accurately express the complex operational constraints of discrete IPs, providing standardized formulations for optimization problems such as production scheduling. Therefore, RTN has been widely applied in various scenarios \cite{castro_resourcetask_2013, zhang_cost-effective_2017, zhang_bidding_2015, su_multi-objective_2023, zhang_demand_2018, caro-ruiz_coordination_2019}. As a typical application, Castro et al.~\cite{castro_resourcetask_2013} used the RTN to model steel-making plants and developed an algorithm to minimize energy costs. Building upon this, Zhang et al.~\cite{zhang_cost-effective_2017} extended the RTN model to incorporate flexible adjustment modes in production processes, which is a crucial source of flexibility for steel plants and aluminum smelters. For example, power consumption in the aluminum smelting process can be changed accurately and quickly by controlling the DC voltage of the electrolyzer without affecting the production quality~\cite{zhang_bidding_2015}.
Recent research has focused on coupling the RTN model with other practical issues in industrial demand response, such as considering the carbon emission intensity of IP~\cite{su_multi-objective_2023}, coordinating IP with energy storage~\cite{zhang_demand_2018}, and integrating renewable energy~\cite{caro-ruiz_coordination_2019}. In the above studies, the RTN model was directly applied or modified based on the energy usage scenarios of discrete IPs. Although the computational efficiency and scalability problems of the RTN model have been noted~\cite{zhang_cost-effective_2017}, they were not the primary focus of previous studies.

In fact, one major challenge of the RTN model lies in its computational complexity. The legacy RTN model requires numerous binary variables to represent the process of resource transformation and task execution in discrete IP operations, leading to poor computational performance of optimization problems (e.g., the production scheduling problem of minimizing industrial energy costs)~\cite{floudas_continuous-time_2004}. These binary variables are necessary for two main reasons: First, they represent the operating states of manufacturing devices that can only switch between several operating points (e.g., ``on'' and ``off'')~\cite{li_real-time_2017}. Second, they model other technical constraints such as the material flow of intermediate products, which must be transported in batches on the production line~\cite{zhang_cost-effective_2017}. The computational burden is further exacerbated because the legacy RTN model treats each batch of products separately, meaning that the number of variables and constraints increases proportionally with production targets. For example, when modeling a steel plant with flexible EAFs using RTN, the model contains over 4000 binary variables for 8 production batches and takes 30 seconds to solve. When the target increases to 12 batches, the binary variables exceed 7000, making the model unsolvable within two hours~\cite{zhang_cost-effective_2017}, thus severely limiting its applicability.

For the extensive development of industrial demand response, the scalability issue of the RTN model should be reasonably addressed, especially in scenarios where many industrial consumers collaborate to optimize their production or industrial consumers' energy consumption constraints need to be embedded into upper-level dispatching problems. Zhang et al.~\cite{zhang_computational_2016} focused on the computational complexity of the production scheduling problem based on the RTN and proposed methods such as adding cuts to make the computations more tractable. However, the modeling techniques for RTNs were not optimized. Nolde et al. proposed a modeling method based on continuous time, but the focus was on the accuracy of following the load, and the scalability of the model was not considered~\cite{nolde_electrical_2010}. Overall, the existing research lacks sufficient attention to the scalability of the RTN model, and to our knowledge, there is almost no research that systematically addresses computational complexity by improving modeling techniques. In our previous work~\cite{lyu_lstn_2023}, considering that state switching time of devices (e.g., 2 minutes) is negligible compared to demand response timescale (e.g., several hours), we modeled the operating duration of the devices within the time intervals as continuous variables, eliminating binary variables in the state-task network model for continuous IPs~\cite{lu_data-driven_2021}. However, as mentioned above, discrete IPs are more complex, and the modeling technique for the state-task network model cannot be directly applied to RTN.

This paper presents a novel approach to modeling discrete IPs for industrial demand response, an effort to systematically improve the computational performance of the legacy RTN model. We change the variable definitions and corresponding constraint formulations based on RTN without affecting the representability of the original model. Specifically, we use continuous variables to represent task operating times and progress in a unified manner for both ordinary and flexible production processes. Based on these variables, we design mathematically favorable forms for technical constraints of discrete IPs, including resource balance, task execution, waiting time, and production objectives. By doing so, we provide a discrete IP model with fewer binary variables, shorter solving time, and better scalability. Finally, we validate the effectiveness of the proposed model through numerical tests, demonstrating improved computational performance for discrete IP models without compromising representability.

The main contributions of this paper are summarized as follows:
\begin{itemize}

  \item We propose a novel continuous resource-task network (cRTN) model that systematically improves the computational efficiency of discrete IP models. The key innovation lies in using continuous variables to represent task progress and resource quantities, while carefully designing constraints to describe the discrete nature of IPs. Our unified modeling approach handles both non-flexible and flexible production processes in a unified manner.
  
  \item We demonstrate through numerical experiments that the cRTN model achieves superior computational performance compared to legacy RTN models. Specifically, the number of binary variables in cRTN remains constant regardless of production targets, while in RTN it scales linearly with the number of batches. This improved scalability is crucial for practical applications such as coordinating multiple IPs or integrating IP constraints into power system dispatch problems.
  
  \item Our reformulation addresses a previously reported but unresolved issue of rounding errors in RTN models. By redesigning the mathematical forms of batch processing and resource transfers, cRTN ensures strict feasibility of discrete IP constraints while enabling more accurate optimization of energy costs. This improvement is validated through case studies showing both reduced computation time and lower operational costs.

\end{itemize}

The remainder of this paper is organized as follows: Section~\ref{sec_problem_description} introduces the characteristics of discrete IPs and our objectives.
Section~\ref{sec_rtn} and Section~\ref{sec_model} present the legacy RTN model and the proposed cRTN model, respectively.
Section~\ref{sec_illustrative} illustrates the modeling process of cRTN compared with the legacy model with an eye-ball example.
Section~\ref{sec_numerical} presents the numerical tests and the results using a steel plant case. Section~\ref{sec_conclusion} concludes the paper and discusses future work.

\section{Problem Description and Methodology}\label{sec_problem_description}
\subsection{Characteristics of discrete production processes}

In typical IPs, raw materials sequentially undergo multiple stages of processing on a production line to become the final product. Compared to continuous IPs, the discrete IPs considered in this paper have the following main characteristics (Table~\ref{tab:ipp_comparison}):

\paragraph{Production that is conducted in batches (Fig.~\ref{fig_discrete_ipp2})} In continuous IPs, the production, storage, consumption, and transportation of materials is a continuous process without minimum processing unit restrictions. For example, a steel powder grinder can process 60kg of steel powder in one hour or 1kg in one minute, naturally lending the product to continuous variables in modeling. In contrast, discrete IPs have minimum processing units (batches) for each task. For instance, an EAF in secondary steel-making processes several tons of scrap steel at once, heating the entire batch until melting. While processing 6 tons per hour, this cannot be modeled as 0.1 tons per minute because the scrap steel is processed as a whole batch - no steel is available for the next process until the full 60-minute EAF operation completes. This characteristic necessitates the introduction of binary variables to describe the production process.

\paragraph{Finite operating states of the production equipment and restrictions on state switching} This common characteristic in IPs typically requires integer variables to describe equipment operating states, similar to the state variables in thermal Unit Commitment (UC). In continuous IPs, although equipment can only operate in discrete states (e.g., on and off), the operating time, energy consumption, and material production in each state are continuously adjustable (e.g., operating for 47 minutes within an hour). This continuous nature formed the basis of our previous work in linearizing continuous IP operation models in DR time scales~\cite{lyu_lstn_2023}. However, this foundation doesn't exist in discrete IPs, where equipment states are not only discrete but also cannot switch freely, being coupled with production progress. For example, once an EAF starts processing a batch of scrap steel, it must complete the batch processing before returning to an idle state and cannot be interrupted midway, similar to the minimum start-up and shut-down time constraints in thermal UC.

\paragraph{Coupling between the preceding and subsequent production stages} For continuous IPs, the coupling between stages is modeled through material transformation matrices~\cite{lyu_lstn_2023, chen_real-time_2024}, and the storage limits for intermediate materials are represented by linear constraints due to the continuous nature of the materials. However, in discrete IPs, due to the batch processing mode, subsequent stages can only begin after the complete completion of previous stages, necessitating additional integer variables. Furthermore, after processing in the prior stage, the product needs to be transferred and may have to wait before being further processed in the subsequent stage. The time for transfer and waiting may be subject to limitations; for example, molten steel from an EAF needs to be processed before it cools.

\subsection{Objectives of Modeling}

\begin{figure}[!t]
  \centering
  \includegraphics[width=3.0in]{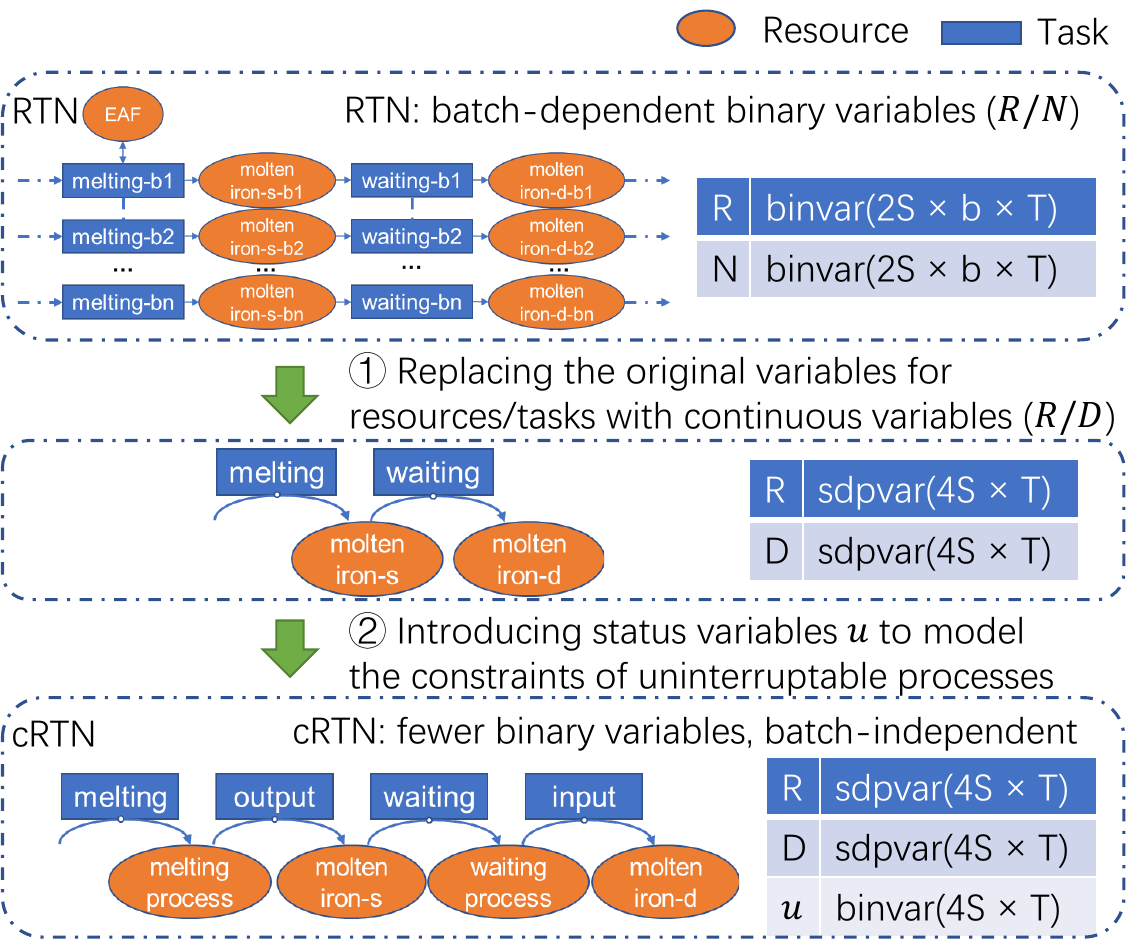}
  \caption{The cRTN model introduces two key improvements over the legacy RTN model: While maintaining the graph-based modeling philosophy where tasks (edges) describe resource (node) conversions, it (1) models resources and task operations using batch-independent continuous variables, replacing the discrete variables in legacy RTN, and (2) introduces status variables to model the constraints of uninterruptible processes, ensuring technical constraints of discrete IPs are strictly satisfied while improving computational efficiency. ($S$: number of stages, $T$: number of time slots, $b$: number of batches)}
  \label{fig_framework}
\end{figure}

\paragraph{Effective Representation of Technical Constraints for Discrete IP} In essence, the model variables will not violate the mathematical constraints proposed by the model if, and only if, the corresponding production scheduling can be executed in reality without infringing upon the technical constraints of the IP. Additionally, we desire the variables within the model to be easily comprehensible so that factory managers can conveniently implement production schedules based on the variable values provided by the model.

\paragraph{Computability and Scalability} The computational complexity introduced by integrating the constraints provided by the proposed model into optimization problems should not be excessively high. Alternatively, it should be solvable within an acceptable time frame through commercial solvers to obtain an optimal solution.

\subsection{Methodology}

Subsequently, we briefly introduce the mathematical forms of the technical constraints under the existing modeling method, and then improve their mathematical forms one by one.
To achieve our objectives, we initially select a state-of-the-art model that can adequately describe the characteristics of typical discrete IPs as mentioned above. The resource-task network (RTN) is a graph-based modeling framework widely used for discrete IP scheduling, where physical elements are unified as resources (nodes) and operations are unified as tasks (edges). Tasks describe the conversion relationships between resources, forming a network structure that represents the entire production process. As previously stated, Zhang et al. incorporated the flexibility of adjusting the power consumption rate of the production processes into the RTN model~\cite{zhang_cost-effective_2017}, a modeling method still widely utilized in recent research~\cite{su_multi-objective_2023}. 

Building upon this foundation, we propose a novel model that inherits RTN's graph-based modeling philosophy while introducing continuous variables to represent resource quantities and task operations. We name our model ``continuous RTN (cRTN)''.
This approach maintains the necessary binary variables only where essential, ensuring accurate representation of both the scheduling process and material flow while improving computational efficiency (Fig.~\ref{fig_framework}). Subsequently, we briefly introduce the mathematical forms of the technical constraints under the existing modeling method~\cite{zhang_cost-effective_2017}, and then improve their mathematical forms one by one.

\section{The Legacy RTN Model}\label{sec_rtn}

In this section, we present the legacy RTN model formulation, which was originally adapted by Zhang et al.~\cite{zhang_cost-effective_2017} from Castro et al.'s work~\cite{castro_resourcetask_2013} and extended to incorporate flexible-adjustable tasks. While we maintain the mathematical essence of their formulation, we have made some adjustments to the notation for consistency with our proposed cRTN model in Section~\ref{sec_model}. This section serves as a foundation for comparison, demonstrating how our cRTN model achieves improved computational efficiency while maintaining the same modeling capabilities.

\subsection{Resource Task Network}

\begin{figure}[!t]
  \centering
  \includegraphics[width=3.5in]{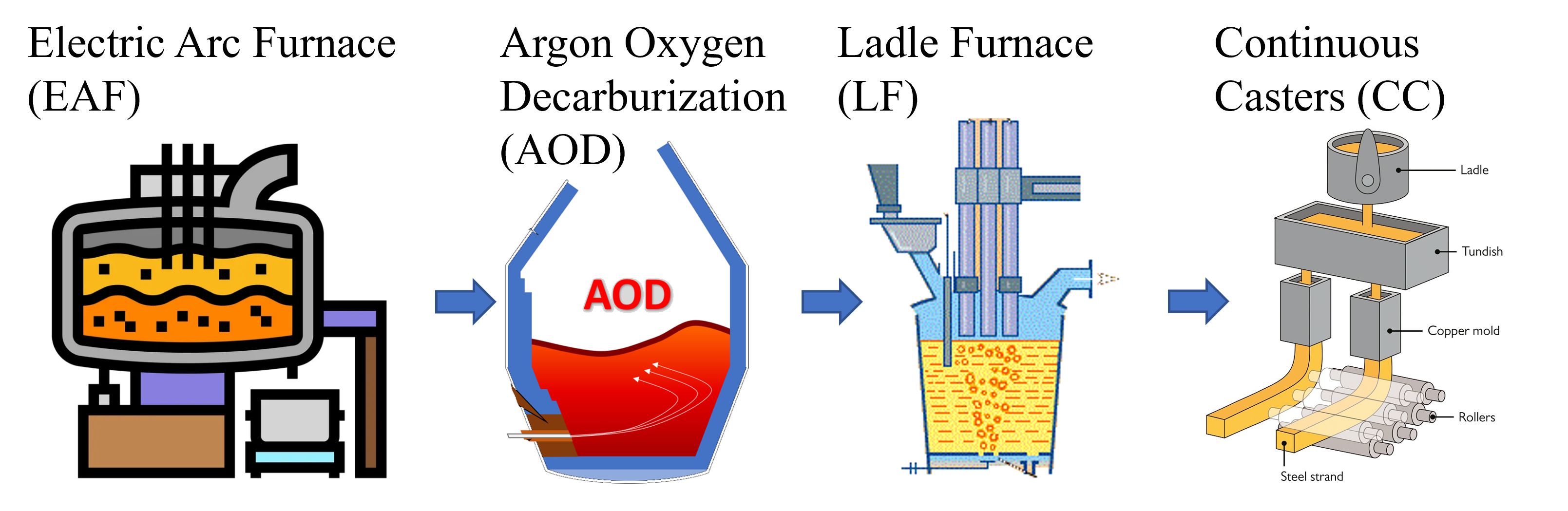}
  \caption{The production process of a steel plant.}
  \label{fig_rtn_process}
\end{figure}

We use the production process of steel manufacturing as an example to illustrate the modeling method of the RTN. The production process of a steel plant is shown in Fig.~\ref{fig_rtn_process}, which can be divided into four stages: 1. melting, 2. decarburization, 3. refining, and 4. casting. The four production stages and the transfer tasks between them are modeled as tasks indexed by $i$. The schedule of tasks is based on discrete time slots indexed by $t$, with a length of $\delta$ (e.g., $\delta=5$ minutes). The binary variable $N_{i, t}$ models the start of task $i$, with $N_{i, t} = 1$ representing that task $i$ starts at time slot $t$.

The production equipment and products (intermediate and final products) are modeled as resources, indexed by $r$. $R_{r, t}$ represents the value of resource $r$ at time $t$. For instance, if $r$ represents the EAF, $R_{r, t} = 1$ means that the EAF is idle at time slot $t$ and can be used for melting. Since the product needs to be transported between different production stages, to distinguish the product before and after transportation, it is modeled as different resources, represented by superscripts s and d, respectively; i.e., $r^{\rm s}$ ($r^{\rm d}$) is the index of the product located at the transfer start point (destination).

The interaction between tasks and resources is modeled by the interaction matrix $M$, in which an entry $\mu_{r, i, \theta}$ represents the amount of resources $r$ consumed/generated by task $i$, $\theta$ time slots after the start of task $i$. For example, the melting task (task $i$) generates molten steel (resource $r$) at 80 minutes (16 time slots with a time slot length of $\delta=5$ minutes) after the starting time, so entry $\mu_{r, i, 16}$ is set to 1.

Note that in the legacy RTN model, tasks are bound to batches; that is, the process and transfer of different batches are modeled as different tasks. Specific indexes, modeling methods, and examples can be found in Ref.~\cite{zhang_cost-effective_2017}. This modeling method means that the scale of the RTN model of the same production line will increase with the production target (the number of batches), limiting the model's scalability.

\subsection{Mathematical Formulation of the Basic RTN Model}

The modeling approaches herein were selected from Ref.~\cite{zhang_cost-effective_2017} and reduced for clarity and brevity.

\paragraph{Resource balance} The amount of resources at the end of each time slot $R_{r, t}$ is determined by the initial amount (i.e., the amount at the end of the previous time slot) $R_{r, t-1}$ and the changes to the resources caused by the tasks within the time slot. This principle of resource balance can be represented as a constraint for both equipment and products in the following form:
\begin{equation}\label{rtn_balance}
  R_{r, t}=R_{r, t-1} + \sum_i \sum_{\theta=0}^{\tau_i} \mu_{r, i, \theta} N_{i, t-\theta} \quad \forall r, t
\end{equation}
Imposing this constraint on the equipment limits the use of a device to a single task within a given time period. However, this leads to rounding errors, which will be further analyzed in the case study. Similarly, the energy consumption of production process $\Pi_{E L, t}$ can be represented with a comparable constraint, although there is no direct coupling between the energy consumption across different time slots:
\begin{equation}\label{rtn_electricity_balance}
  \Pi_{E L, t}=\sum_i \sum_{\theta=0}^{\tau_i} \mu_{E L, i, \theta} N_{i, t-\theta} \quad \forall t
\end{equation}
where $\mu_{E L, i, \theta}$ is the energy consumption of task $i$, $\theta$ time slots after the task starts.

\paragraph{Task Execution} The RTN models each batch of products individually, and task modeling is also based on batches. Therefore, each batch needs to have each task applied to it only once:
\begin{equation}\label{rtn_task_execution}
  \sum_{t} N_{i, t} = 1 \quad \forall i
\end{equation}
For transfer tasks, the constraint below is added to ensure immediate execution (\ref{rtn_transfer_exc}). Constraint (\ref{rtn_transfer_exc}) requires no waiting time for intermediate products at the start of the transfer. The rationale behind this is twofold: first, this is a common requirement in IPs such as steel manufacturing; second, it avoids a modeling approach where manufacturing is completed in one stage, followed by a waiting period, followed by transfer, followed by waiting again; avoiding this approach maintains the model's simplicity.
\begin{equation}\label{rtn_transfer_exc}
  R_{r^{\rm s}, t} = 0 \quad \forall r^{\rm s}, t
\end{equation}

\paragraph{Waiting Time Limit} Intermediate products can wait for some time between two production stages rather than immediately proceeding to the next stage. This time flexibility is one source of the energy flexibility of IPs. However, this waiting time usually has an upper limit because materials such as molten steel must be further processed before their temperature drops to a certain threshold. The duration limit of the waiting process (including the transfer time) is expressed as follows
for resource $r$:
\begin{equation}\label{rtn_transfer_time}
  \delta \sum_t R_{r^{\rm d}, t} + w_{r^{\rm d}} \le W_{r^{\rm d}} \quad \forall i
\end{equation}
where $R_{r^{\rm d}, t} = 1$ indicates that the product has arrived at the transfer destination at time slot $t$ and is waiting; therefore, $\sum_t R_{r^{\rm d}, t}$ is the waiting time, $w_{r^{\rm d}}$ is the transfer time and $W_{r^{\rm d}}$ is the maximum waiting time of resource $r^{\rm d}$.

\paragraph{Product Delivery} By the end of the final time slot, each batch needs to reach the final stage, i.e., complete all the manufacturing processes:
\begin{equation}\label{rtn_product_delivery}
  R_{r^{rm f}, T} = 1, \forall r^{rm f}
\end{equation}
where $r^{rm f}$ is the index of the product located at the final stage.

\paragraph{Objective} The goal of factory production scheduling is not a constraint on its internal production process. For the completeness of the model, we use an objective function as an example, which aims to minimize the production energy cost:
\begin{equation}\label{rtn_objective}
  {\rm min.} \sum_{t} Pr_{t}  \Pi_{E L, t}
\end{equation}
where $Pr_{t}$ is the electricity price at time slot $t$.

\paragraph{Multiple operating modes} In actual operation, a device may operate in one of several states, not just in the ``on'' and ``off'' states. The basic RTN model mentioned above can model this feature by modeling the different operating states of the device as different tasks. For example, suppose a device can operate at 50\% of its rated power. In that case, the interaction matrix needs to model its processing time as twice the rated operating state (i.e., $\mu_{r, i, 2\tau} = 1$, where $\tau$ is the nominal processing time), and the energy consumption per time slot is half the nominal value.

\subsection{Flexible operating mode}
If a device can adjust its operating point in each period or even continuously adjust within the operating boundaries, then its processing time can be adjusted within a feasible range. This constraint can be expressed in terms of the start time of the subsequent task (i.e., the end time of the task, as we assume the transfer tasks are executed immediately) as follows:
\begin{equation}\label{rtn_flexible_mode_1}
  \sum_{t' = t+\tau^L_i}^{t+\tau^H_i} N_{i+, t'} \ge N_{i, t}
\end{equation}
where $\tau^L_i$ and $\tau^H_i$ are the minimum and maximum processing times of task $i$, respectively, and $i$ and $i+$ are the indices of the task that can be flexibly adjusted and its subsequent task, respectively.

Under the flexible operating mode, we also need to introduce the power variable of the device operation (a continuous variable), which is subject to the following constraints:
\begin{equation}\label{rtn_flexible_mode_2}
  P^{L}_{i} \cdot S_{i, t} \le P_{i, t} \le P^{H}_{i} \cdot S_{i, t} \quad \forall i, t
\end{equation}
where $P^{L}_{i}$ and $P^{H}_{i}$ are the minimum and maximum processing powers of task $i$, respectively; $P_{i, t}$ is the power of task $i$ at time slot $t$; $S_{i, t}$ is the processing status of task $i$ at time slot $t$; and $S_{i, t} = 1$ indicates that task $i$ is being processed at time slot $t$, which means that the device is on and usually cannot be interrupted until the end of the task. Naturally, $S_{i, t}$ is determined by the start times of the task and the subsequent task:
\begin{equation}\label{rtn_flexible_mode_3}
  S_{i, t} - S_{i, t-1} = N_{i, t} - N_{i+, t} \quad \forall t
\end{equation}

Finally, different control methods need to comply with the energy-material conversion conditions of processing itself, and a certain amount of energy needs to be invested to complete processing:
\begin{equation}\label{rtn_energy_requirement}
  \sum_t \delta P_{i, t} \ge E^{\rm req}_i \quad \forall i
\end{equation}
where $E^{\rm req}_i$ is the energy required to complete task $i$, which can be determined by the nominal power multiplied by the nominal processing time of the task if it is assumed that the same amount of energy is required for different processing modes.

\section{Continuous RTN Formulation}\label{sec_model}

While maintaining the same representational capabilities as the legacy RTN model presented in Section~\ref{sec_rtn}, our cRTN model introduces two fundamental improvements. First, the modeling of tasks and resources is decoupled from batches. In other words, the same product or process in the same stage is modeled with only one resource or task, respectively. Thus, the scale of the model will not increase with the production target, as in legacy RTNs.

Second, the decision variables of the model are changed from the starting time slot, operating states, and power at each time slot to the operating time for each state within each time slot. This idea is consistent with our previous work~\cite{lyu_lstn_2023}. It utilizes the relatively large time scale of demand response to make the decision variables continuous, partly sacrificing model accuracy to obtain better computational performance. The specific mathematical form of the continuous RTN model is given below.

\subsection{Mathematical Formulation of the Continuous RTN Model}

In the following model, $i$ and $r$ represent the task and resource indexes, respectively. We still use $R_{r, t}$ to represent the quantity of resources, which is relaxed to a continuous variable. $R_{r, t}$ can be understood as the progress of the task that generates resource $r$. $R_{r, t}=0(1)$ represents no progress (task completion), and $R_{r, t} \in (0, 1)$ represents that processing is ongoing. This design enables our model to represent ordinary processing procedures and flexible (power-adjustable) production processes in a unified form, avoiding the separate modeling required in legacy RTNs.

\begin{figure}[!t]
  \centering
  \includegraphics[width=3.0in]{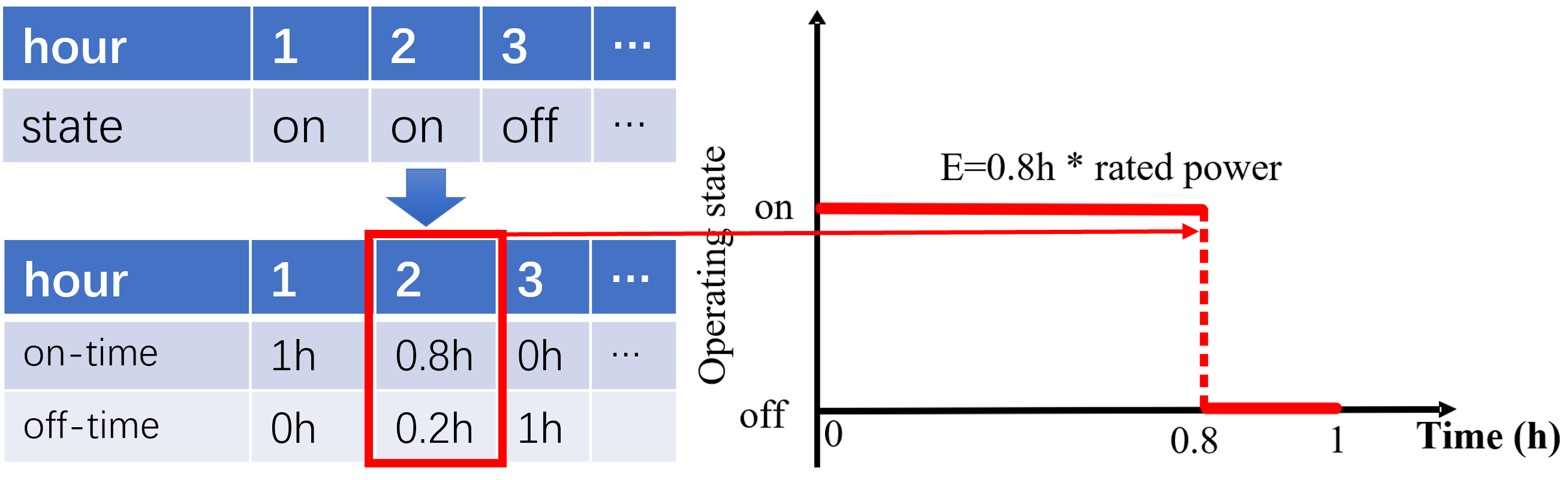}
  \caption{The continuous RTN no longer uses the on/off states of the device as variables but instead uses the duration in which the device operates in various states within the time slots.}
  \label{fig_continous_time}
\end{figure}

The continuous variable $D_{i, k, t}$ represents the time that task $i$ runs in state $k$ within time slot $t$ (Fig.~\ref{fig_continous_time}).
The task-resource association matrix $G$ is introduced to represent the resource change for each task.
The matrix entry $g_{r,i,k}$ represents the changed amount (generated or consumed) of resource $r$ when task $i$ operates in state $k$ per unit of time.

For ease of description, we assume by default that the state $k=0$ corresponds to the idle state of the task ($g_{r, i, 0}=0$), and $k=1$ and $k=2$ represent the states corresponding to the minimum and maximum processing speeds after the task starts running, respectively. For example, for the melting process, under the flexible adjustment mode, the processing time can be controlled to between 75 and 125 minutes. Hence, the generation rate, i.e., the amount of product (in batches) generated per unit time, is $g_{r, i, 1}=1/125$ to $g_{r, i, 2}=1/75$ (per minute). For tasks with only one processing speed, it is only necessary to model two states: $k=0$ (off) and $k=1$ (on). This design allows our model to uniformly represent regular and flexible production processes instead of requiring separate modeling, as in legacy RTNs.

\paragraph{Resource balance}
Using continuous decision variables of tasks, the resource changes by task are also continuous. The resource balance constraint can be expressed by (\ref{crtn_balance_1}):
\begin{equation}\label{crtn_balance_1}
  R_{r, t} \le R_{r, t-1} + \sum_{i} \sum_{k} g_{r, i, k} D_{i, k, t} \quad \forall r, t.
\end{equation}
Note that we use ``$\le$'' for two reasons. On the one hand, this approach can avoid the problem of conflict with other constraints for modeling based on discrete time slots (e.g., Equation (\ref{crtn_r_limit})), which was also reported by Zhang et al.~\cite{zhang_cost-effective_2017} but has been neglected. Using ``$\le$'' does not change the effect of the constraint since equality must be enforced when minimizing the energy cost. On the other hand, it is possible to satisfy ``$<$'' by idling the equipment for a short time.
To validate this relaxation, we analyzed the gap between its two sides under different optimality gap settings (see Fig.~\ref{fig_resource_balance_gap} in the case study), demonstrating that our relaxation does not compromise the model's accuracy.

A task can only process products in batches, and usually, multiple batches will not be allowed to accumulate at the same stage; therefore, we have:
\begin{equation}\label{crtn_r_limit}
  0 \le R_{i, t} \le 1 \quad \forall i, t.
\end{equation}

\paragraph{Task Execution}
By definition, $D_{i, k, t}$ is non-negative and does not exceed the length of the time interval (\ref{crtn_task_execution_1}), and the total operating time of task $i$ in time slot $t$ is equal to the length of the time interval (\ref{crtn_task_execution_2}):
\begin{equation}\label{crtn_task_execution_1}
  0 \le D_{i, k, t} \le \delta \quad \forall i, t
\end{equation}
\begin{equation}\label{crtn_task_execution_2}
  \sum_{k} D_{i, k, t} = \delta \quad \forall i, t
\end{equation}
This formulation is quite different from the legacy RTN model, where the task execution is modeled as a binary variable. The formulation is inspired by our previous work on continuous IPs~\cite{lyu_lstn_2023}, where the task execution is modeled as a continuous variable considering the context of demand response (i.e., the time scale of demand response in hours is much larger than the decision time step of the production process in minutes and the task execution duration can be adjusted continuously by determining the start and/or end time of the task within the time slot). By using this formulation, we build a continuous RTN model on the basis of the linearized version of the state-task network (STN) model, indicating that continuous production processes are a degenerated case of discrete production processes (see Appendix~\ref{app_stn_crtn_comparison}). In other words, we can build discrete IP models by adding constraints to continuous IP models that describe the discrete nature of IPs.

\paragraph{Product Delivery} The amount of the final product $i^{\rm end}$ needs to reach the target by the end of the ending time slot, which can be expressed as:
\begin{equation}\label{crtn_product_delivery}
  R_{i^{\rm end}, T} \ge R^{\rm tg}_{i^{\rm end}}
\end{equation}
where $R^{\rm tg}_{i^{\rm end}}$ is the target amount of product at the final stage.

\paragraph{Objective} The objective function is the same as that of the legacy RTN model:
\begin{equation}\label{crtn_objective}
  {\rm min.} \sum_{hr} Pr_{hr} \sum_{t \in T_{hr}} \Pi_{E L, t}
\end{equation}
Let $P_{i, k}$ represent the power consumption of task $i$ operating in state $k$; then, the energy consumption of the whole production process in time slot $t$ is given by:
\begin{equation}\label{crtn_electricity_balance}
  \Pi_{E L, t} = \sum_i \sum_{k} P_{i, k} D_{i, k, t} \quad \forall t
\end{equation}

\subsection{Modeling Discrete IPs}

\begin{figure}[!t]
  \centering
  \includegraphics[width=3.49in]{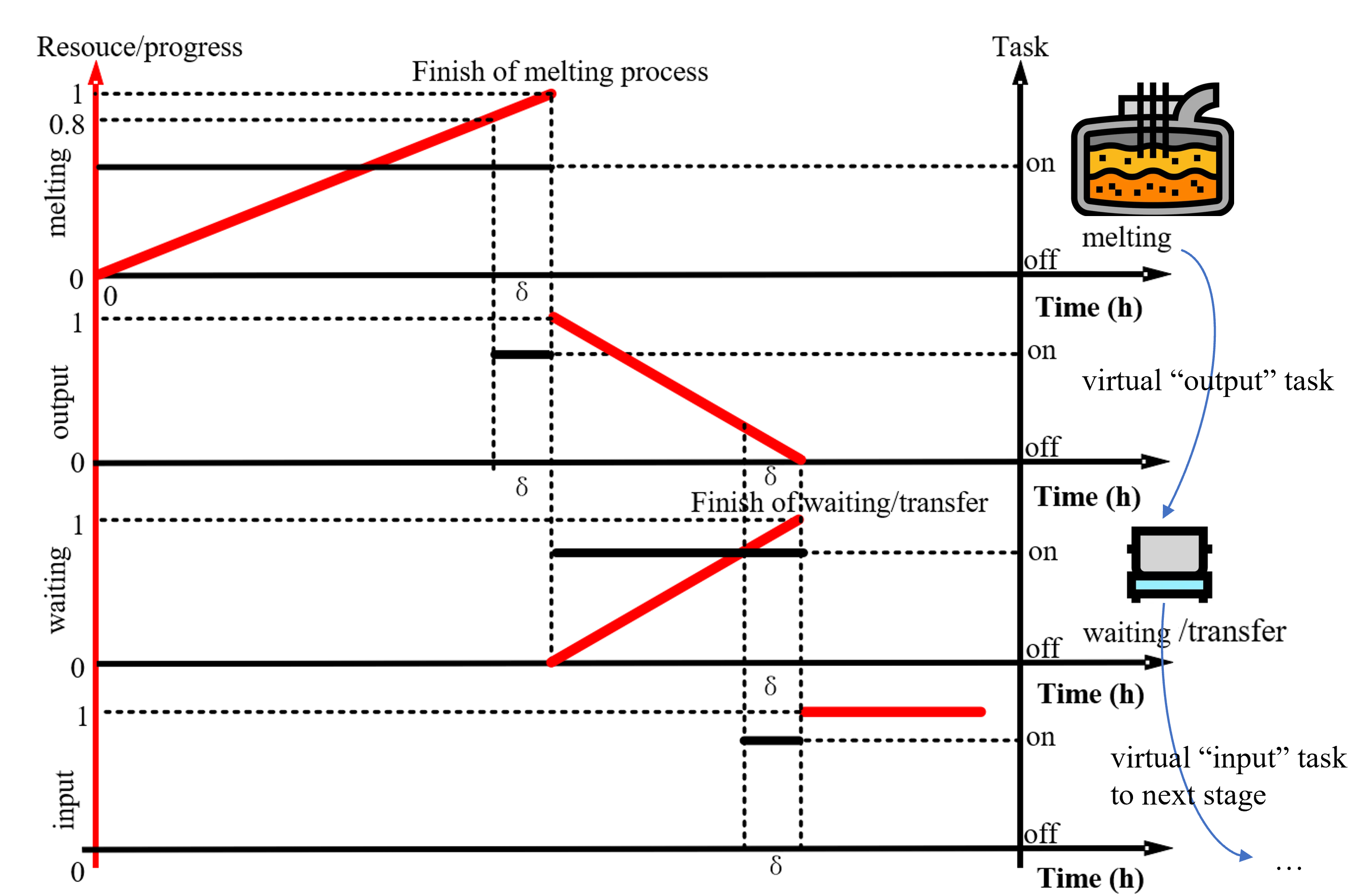}
  \caption{We add ``virtual'' output and input tasks between the ``real'' process and waiting/transfer tasks. At the time slot when the preceding task is completed, the output/input tasks set their resources to 0 and the subsequent task resources to 1. This design expresses the batch processing mode and the sequential relationship of the discrete production process while maintaining the simplicity of the model.}
  \label{fig_process}
\end{figure}

Compared to legacy RTN models, the task characteristics of discrete IPs are modeled differently by the proposed model. It's important to note that our approach is a reformulation rather than a relaxation of the RTN model - while we use continuous variables to represent resources and task progress, we carefully design constraints to preserve the discrete nature of IPs. The key modeling aspects are:

\paragraph{Uninterruptible tasks} Once some tasks start running, they cannot stop until the batch is processed. This type of task (such as the melting task) needs to satisfy $R_{r^{i+}, t}D_{i, 0, t} = 0$, where $r^{i+}$ is the resource produced by task $i$. It is easy to verify that when $R_{r^{i+}, t}>0$, $D_{i, 0, t}=0$; that is, the task producing $R_{r^{i+}, t}$ has started, and it cannot be in the idle state ($k=0$), so it can only operate between the minimum ($k=1$) and maximum ($k=2$) processing speeds. To avoid bilinear terms, we introduce binary variables $u_{i, t}$ to characterize the processing state of the task, where $u_{i, t}=1$ represents that the current process is ongoing.
For uninterruptible tasks, the above constraint can be rewritten as:
\begin{equation}\label{crtn_task_execution_3}
  D_{i, 0, t}  \le 1 - u_{i, t} \quad \forall t
\end{equation}
\begin{equation}\label{crtn_task_execution_4}
  R_{r^{i+}, t} \le u_{i, t} \quad \forall t
\end{equation}
These constraints mathematically ensure that once $R>0$, the task cannot be interrupted, preserving the discrete nature of the original constraints.

\paragraph{Batch-based processing} The subsequent task that consumes resource $r$ can only be carried out when the previous task is completely processed (i.e., $R_{r, t}=1$). To reflect this characteristic of discrete IPs, we add ``output'' tasks and ``input'' tasks between waiting tasks and processing tasks (Fig.~\ref{fig_process}). An output (input) task remains idle until the resource it consumes reaches 1; it is set to 0 in the same time slot when the previous task is completed, while the resource it generates is set to 1, thus meeting the discreteness requirement of batch processing. After introducing $u_{i, t}$, this characteristic of output (input) tasks can be expressed by:
\begin{equation}\label{crtn_task_execution_5}
  D_{i, 1, t} = u_{i, t} \delta \quad \forall t
\end{equation}
For resource $r^{i-}$ that the output (input) tasks consume, set $g_{r^{i-},i,1} = - 1/\delta$; for resource $r^{i+}$ that the output (input) tasks generate, set $g_{r^{i+},i,1} = 1/\delta$. It is easy to verify that under constraints (\ref{crtn_balance_1}-\ref{crtn_task_execution_4}), (\ref{crtn_task_execution_5}) naturally meets the above requirements.

\paragraph{Waiting time limit} We model the waiting times between the production stages of intermediate products as tasks. The waiting tasks consume resources passed on by the preceding output tasks and generate resources for the input tasks. Waiting itself does not generate resources, so the resources here only represent the progress of waiting, and the parameters of the consumption (generation) rate are set to ensure that the total time for transferring and waiting for the intermediate products meets the requirements. Therefore, the minimum and maximum generation rates of the waiting tasks are set to $1/(w_i+W_i)$ and $1/w_i$, respectively, where $w_i$ and $W_i$ are the transfer time and maximum waiting time of task $i$, respectively. Naturally, the waiting tasks are also uninterruptible (satisfying Equations (\ref{crtn_task_execution_3}-\ref{crtn_task_execution_4})) because once the product reaches this task, waiting cannot be ``interrupted''.

Finally, since the waiting tasks already consider the waiting time, other tasks must start running immediately when their input materials arrive and cannot remain idle. This can be represented by $R_{r^{i-}, t}D_{i, 0, t}=0$, where $r^{i-}$ represents the resources consumed by task $i$. Similarly, this constraint can be replaced by (\ref{crtn_task_execution_3}) and (\ref{crtn_task_execution_6}):
\begin{equation}\label{crtn_task_execution_6}
  R_{r^{i-}, t} \le u_{i, t} \quad \forall t
\end{equation}

It should be noted that this section presents only the fundamental principles and illustrations of the modeling approach. In practice, production scheduling problems with finite time horizons require additional constraints for the initial and final time slots. For example, constraints must be added to prevent tasks with durations exceeding two time slots from starting in the final time slot. For detailed implementation of these specific cases, readers can refer to our code implementation at \cite{rick10119_crtn_2024}.

\section{An Illustrative Example}\label{sec_illustrative}

\begin{figure}[!t]
  \centering
  \includegraphics[width=2.0in]{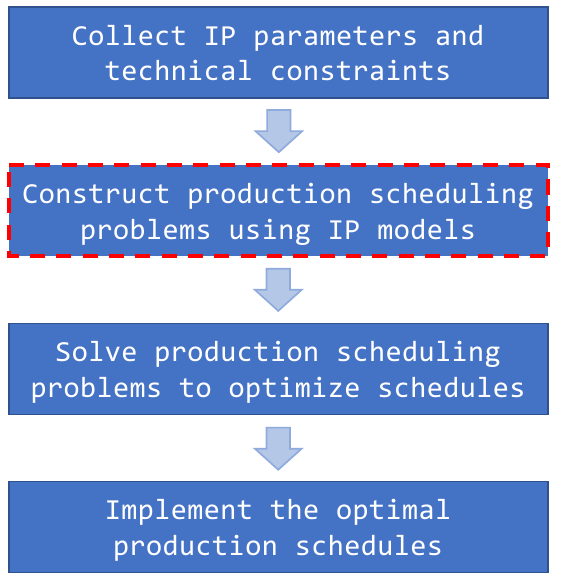}
  \caption{Flowchart of the IP scheduling process, consisting of four main steps: (1) collecting IP parameters from equipment specifications and expert knowledge, (2) constructing the optimization model using either RTN or proposed cRTN formulation, (3) solving the model using commercial solvers, and (4) transforming the optimization results into executable operations. Step 2 represents the key contribution of this paper - the model transformation from discrete to continuous variables while maintaining strict feasibility of discrete IP constraints.}
  \label{fig_flowchart}
\end{figure}

To illustrate the conceptual and mathematical differences between RTN and cRTN modeling approaches, we present a simple example of modeling the same discrete IP using both methods. Specifically, we model an electric arc furnace (EAF) in a secondary steelmaking plant to minimize electricity costs while satisfying technical constraints and production targets. The EAF processes scrap steel in batches through heating and melting. Its original parameters are shown in Table~\ref{tab_parameter_illustrative}, where time parameters are expressed in terms of time slots with a slot length of 1 hour. We consider a 6-hour scheduling horizon, resulting in a single production stage ($S=1$) and time horizon $T=6$. The production target is to provide 2 batches of molten iron ($b=2$) for subsequent production stages by the end of the time horizon. Time-of-use electricity prices are presented in Table~\ref{tab_price_scenarios}. The flowchart of the IP scheduling process is shown in Fig.~\ref{fig_flowchart}, where Step 2-construction of the production scheduling problem is where the RTN or cRTN model is applied.

\begin{table}[!t]
  \caption{Parameters of the electric arc furnace.}
  \label{tab_parameter_illustrative}
  \centering
  \begin{tabular}{ll}
    \toprule
    parameter                   & value \\ \midrule
    nominal power (MW)        & 1  \\ 
    processing time (minute/time slot)  & 120/2  \\
    transfer time (minute/time slot)    & 60/1  \\
    max waiting time (minute/time slot) & 120/2 \\ \bottomrule
  \end{tabular}
\end{table}

\begin{table}[!t]
  \caption{Time-of-use electricity prices for two scenarios (\$/MWh).}
  \label{tab_price_scenarios}
  \centering
  \begin{tabular}{lcccccc}
    \toprule
    Hour & 1 & 2 & 3 & 4 & 5 & 6 \\ \midrule
    Scenario 1 & 100 & 100 & 100 & 100 & 200 & 200 \\
    Scenario 2 & 100 & 100 & 100 & 200 & 200 & 100 \\ \bottomrule
  \end{tabular}
\end{table}

\subsection{Conventional RTN Modeling}

\begin{figure}[!t]
  \centering
  \includegraphics[width=3.0in]{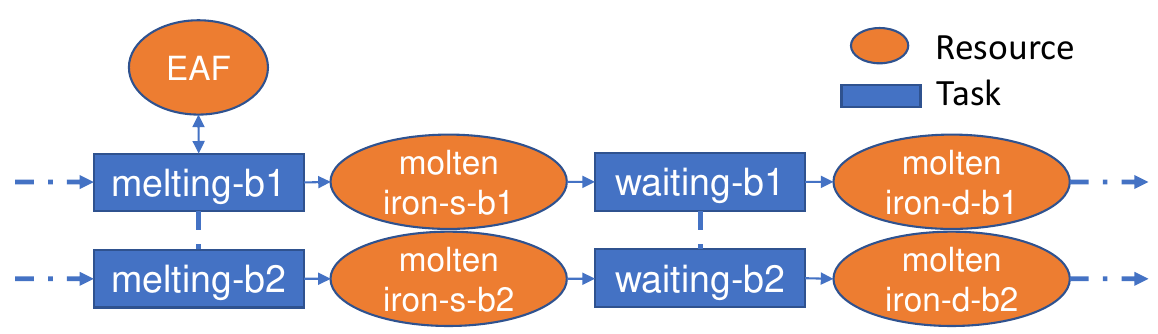}
  \caption{The RTN model for the EAF.}
  \label{fig_rtn_model}
\end{figure}

In the RTN framework (Fig.~\ref{fig_rtn_model}), EAF-related tasks include: 1. Processing (melting), 2. transferring, and 3. Waiting. To simplify the model, transferring and waiting tasks can be combined into a single waiting task. This results in the RTN requiring modeling of 5 types of resources and 4 types of tasks, as listed in Tables~\ref{tab_rtn_resources} and~\ref{tab_rtn_tasks}, where b1 and b2 represent the two batches. The number of resources, tasks, and corresponding variables scales with the number of batches (Table~\ref{tab_rtn_variables}). Table~\ref{tab_rtn_tasks} also provides examples of vectors in the interaction matrix modeling task-resource relationships. For instance, $\mu_{1,1,[\theta]}= [-1, 0, 1]$ indicates that the melting task for batch 1 occupies the EAF at its start time slot and releases it after two time slots. Similarly, $\mu_{2,1,[\theta]}= [0, 0, 1]$ shows that the melting task for batch 1 produces one unit of Molten iron-s-1 two time slots after initiation, where s(d) denotes the product located at the transfer start (destination).

\begin{table}[!t]
  \caption{Tasks and their resource interactions in the RTN model.}
  \label{tab_rtn_tasks}
  \centering
  \begin{tabular}{llll}
    \toprule
    Index (i) & Task & \begin{tabular}[c]{@{}l@{}}Related\\ resources (r)\end{tabular} & \begin{tabular}[c]{@{}l@{}}Interaction matrix\\ $\mu_{r,i,[\theta]}$\end{tabular} \\ 
    \midrule
    1 & melting-b1 & 1, 2 & \begin{tabular}[c]{@{}l@{}}$[-1, 0, 1]$, $[0, 0, 1]$\end{tabular} \\
    2 & melting-b2 & 1, 3 & \begin{tabular}[c]{@{}l@{}}$[-1, 0, 1]$, $[0, 0, 1]$\end{tabular} \\
    3 & waiting-b1 & 2, 4 & \begin{tabular}[c]{@{}l@{}}$[-1, 0]$, $[0, 1]$\end{tabular} \\
    4 & waiting-b2 & 3, 5 & \begin{tabular}[c]{@{}l@{}}$[-1, 0]$, $[0, 1]$\end{tabular} \\
    \bottomrule
  \end{tabular}
\end{table}

\begin{table}[!t]
  \caption{Resources and their task associations in the RTN model.}
  \label{tab_rtn_resources}
  \centering
  \begin{tabular}{lll}
    \toprule
    Index (r) & Resource & Related tasks (i) \\ 
    \midrule
    1 & EAF & 1, 2 \\
    2 & molten iron-s-b1 & 3, 4 \\
    3 & molten iron-s-b2 & 3, 4 \\
    4 & molten iron-d-b1 & 3, 4 \\
    5 & molten iron-d-b2 & 3, 4 \\
    \bottomrule
  \end{tabular}
\end{table}

\begin{table}[!t]
  \caption{Variables in the RTN model.}
  \label{tab_rtn_variables}
  \centering
  \begin{tabular}{lll}
    \toprule
    Variable & Type & Number \\ 
    \midrule
    $R$: value of the resources & binary & $S(1+2b)T=30$ \\
    $N$: start time of the tasks & binary & $S(2b)T=24$ \\
    \bottomrule
  \end{tabular}
\end{table}

Under this model, the RTN constraints are given by Equations (\ref{rtn_balance})-(\ref{rtn_product_delivery}), while the plant's electricity cost is represented by Equation (\ref{rtn_objective}), which calculates the total electricity procurement cost under the given time-of-use prices (Table~\ref{tab_price_scenarios}). The production scheduling problem based on this RTN formulation is an integer programming problem. We denote the scheduling results from this model as the Basic RTN solution.

For flexible tasks, additional power variables $P$ must be introduced to model the adjustable range (we assume power can be adjusted between 1/2 and 4/3 of nominal value). This introduces new continuous variables and additional constraints (Equations (\ref{rtn_flexible_mode_1})-(\ref{rtn_energy_requirement})), transforming the original integer programming problem into a mixed-integer linear programming (MILP) problem. The RTN modeling approach suffers from poor scalability due to: 1) the large number of binary variables, and 2) the problem size scaling approximately linearly with the number of batches. These limitations motivate our development of an alternative modeling approach.

\subsection{Continuous RTN Modeling}

\begin{table}[!t]
  \renewcommand{\arraystretch}{1.0}
  \caption{Parameters of the cRTN Model for the EAF.}
  \label{tab_model_parameter_illustrative}
  \centering
  \begin{tabular}{lllll}
  \toprule
  task (index)                    & \begin{tabular}[c]{@{}l@{}}operating\\ state (k)\end{tabular} & \begin{tabular}[c]{@{}l@{}}processing\\ rate (/$\delta$)\end{tabular} & \begin{tabular}[c]{@{}l@{}}operating\\ power (MW)\end{tabular} \\
  \midrule
  melting (1) & 0                          & 0                          & 0                          \\
                 & 1                          & $1/4$              & $1/4$                  \\
                 & 2                          & $2/3$             & $2/3$                 \\ \hline
  output (2)     & 0                          & 0                          & 0                          \\
                 & 1                          & 1                          & 0                          \\ \hline
  waiting (3)    & 0                          & 0                          & 0                          \\
                 & 1                          & 1/3                      & 0                          \\
                 & 2                          & 1/2                        & 0                          \\ \hline
  input (4)      & 0                          & 0                          & 0                          \\
                 & 1                          & 1                          & 0                          \\  \bottomrule
\end{tabular}
\end{table}

\begin{figure}[!t]
  \centering
  \includegraphics[width=3.0in]{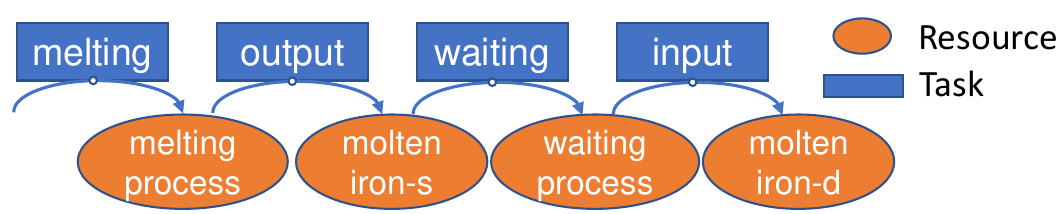}
  \caption{The cRTN model for the EAF.}
  \label{fig_cRTN_model}
\end{figure}

In the cRTN model (Fig.~\ref{fig_cRTN_model}), we similarly combine the transferring and waiting tasks to simplify the model. Thus, the parameter of the combined task (referred to as waiting) is set to 1/3 and 1/2 for the minimum and maximum power, respectively (Table~\ref{tab_model_parameter_illustrative}).
The cRTN requires modeling of 4 types of resources and 4 types of tasks, as shown in Tables~\ref{tab_crtn_tasks} and~\ref{tab_crtn_resources}. Notably, unlike the legacy RTN, the number of resources, tasks, and corresponding variables is independent of the number of batches (Table~\ref{tab_crtn_variables}). The association matrix entries $g_{r,i,k}$ represent the rate of resource generation (negative values indicate consumption). For the EAF with flexible power adjustment between 1/4 and 2/3 of nominal power, the modeling approach remains consistent with non-flexible tasks: $g_{1,1,[k]}= [0, 1/4, 2/3]$ represents the progress rate of the melting task per time slot in states 0 (idle), 1 (minimum power), and 2 (maximum power), respectively.

\begin{table}[!t]
  \caption{Tasks and their resource associations in the cRTN model.}
  \label{tab_crtn_tasks}
  \centering
  \begin{tabular}{llll}
    \toprule
    Index (i) & Task & \begin{tabular}[c]{@{}l@{}}Related\\ resources (r)\end{tabular} & \begin{tabular}[c]{@{}l@{}}Association matrix\\ $g_{r,i,[k]}$\end{tabular} \\ 
    \midrule
    1 & melting & 1 & $[0, 1/4, 2/3]$ \\
    2 & output & 1, 2 & \begin{tabular}[c]{@{}l@{}}$-[0, 1]$, $[0, 1]$\end{tabular} \\
    3 & waiting & 2, 3 & \begin{tabular}[c]{@{}l@{}}$-[0, 1/3, 1/2]$, $[0, 1/3, 1/2]$\end{tabular} \\
    4 & input & 3, 4 & \begin{tabular}[c]{@{}l@{}}$-[0, 1]$, $[0, 1]$\end{tabular} \\
    \bottomrule
  \end{tabular}
\end{table}

\begin{table}[!t]
  \caption{Resources and their task associations in the cRTN model.}
  \label{tab_crtn_resources}
  \centering
  \begin{tabular}{llll}
    \toprule
    Index (r) & Resource & Generated by task & Consumed by task \\ 
    \midrule
    1 & melting process & 1 & 2 \\
    2 & molten iron-s & 2 & 3 \\
    3 & waiting process & 3 & 4 \\
    4 & molten iron-d-1 & 4 & next stage \\
    \bottomrule
  \end{tabular}
\end{table}

\begin{table}[!t]
  \caption{Variables in the cRTN model.}
  \label{tab_crtn_variables}
  \centering
  \begin{tabular}{lll}
    \toprule
    Variable & Type & Number \\ 
    \midrule
    $R$: value of the resources  & continuous & $S(4)T=12$  \\
    $D$: operating duration of the resources & continuous & $S(3*4)T=36$ \\
    $u$: operating status of the tasks & binary & $S(4)T=24$ \\
    \bottomrule
  \end{tabular}
\end{table}

Under this formulation, the production scheduling model based on cRTN (Equations (\ref{crtn_balance_1})-(\ref{crtn_task_execution_6})) is an MILP problem. Since not all processes on the production line are necessarily uninterruptible, the cRTN model requires at most $4ST$ binary variables, independent of the number of batches. In contrast, the RTN model requires $4bST$ binary variables, which scales linearly with the number of batches $b$. For larger-scale problems, such as when $b=10$, the number of binary variables in the RTN model would be ten times that of the cRTN model.

\subsection{Comparison of scheduling results}

\begin{figure}[!t]
  \centering
  \includegraphics[width=3.49in]{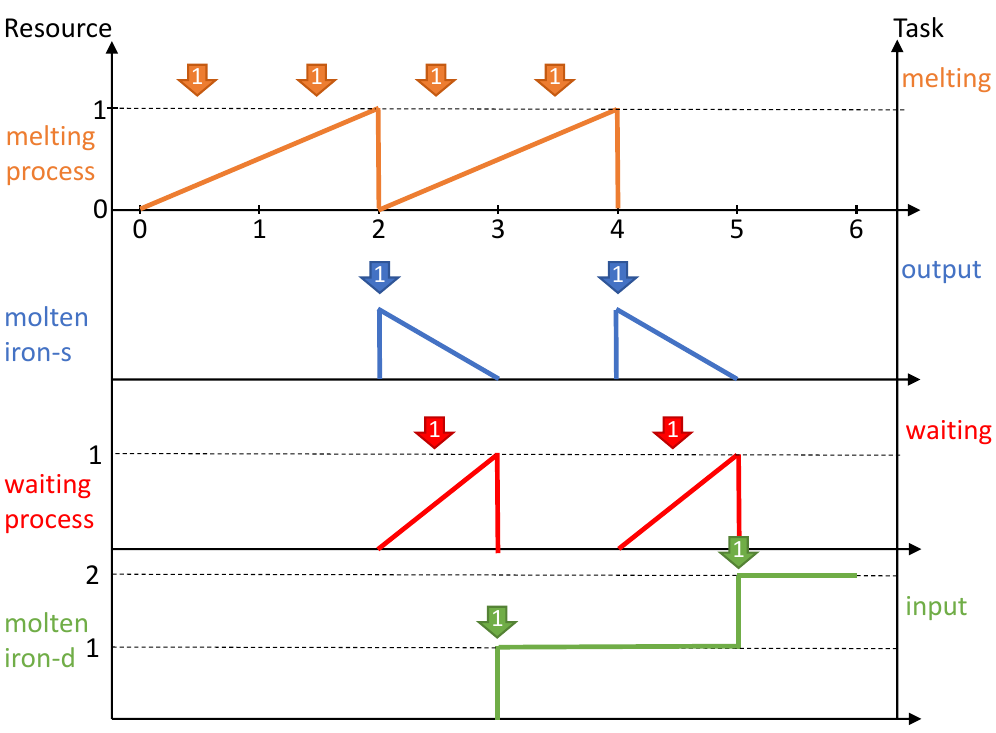}
  \caption{The scheduling results of cRTN in scenario 1.}
  \label{fig_cRTN_result_scenario1}
\end{figure}

\begin{figure}[!t]
  \centering
  \includegraphics[width=3.49in]{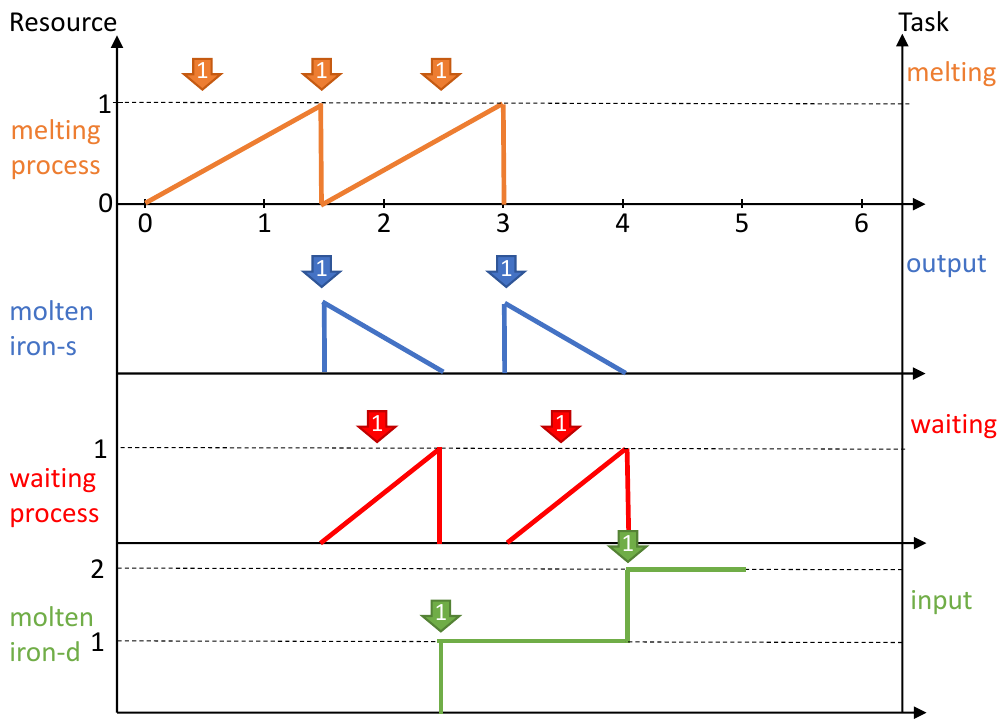}
  \caption{The scheduling results of cRTN in scenario 2.}
  \label{fig_cRTN_result_scenario2}
\end{figure}

We compared the production scheduling results based on RTN and cRTN under two electricity price scenarios, with the main decision variables and production status listed in Tables \ref{tab:scenario1} and \ref{tab:scenario2} (values at the end of each time slot). For RTN, the earliest start time N can be at the end of time slot 0 (beginning of time slot 1) to complete the production of 1 batch by the end of time slot 2. The cRTN scheduling results are also visualized in Figs. \ref{fig_cRTN_result_scenario1} and \ref{fig_cRTN_result_scenario2}.

\begin{table}[!t]
  \caption{Comparison of RTN and cRTN Scheduling Results in Scenario 1.}
  \label{tab:scenario1}
  \centering
  \begin{tabular}{llcccccc}
    \toprule
    Model & Time slot & 1 & 2 & 3 & 4 & 5 & 6 \\
    \midrule
    \multirow{4}{*}{RTN} 
    & EAF power/MW & 1 & 1 & 1 & 1 & 0 & 0 \\
    & molten iron-s & 0 & 1 & 0 & 1 & 0 & 0 \\
    & waiting & 0 & 1 & 0 & 1 & 0 & 0 \\
    & molten iron-d & 0 & 0 & 1 & 1 & 2 & 2 \\
    \midrule
    \multirow{4}{*}{cRTN}
    & EAF power/MW & 1 & 1 & 1 & 1 & 0 & 0 \\
    & molten iron-s & 0 & 1 & 0 & 1 & 0 & 0 \\
    & waiting & 0 & 1 & 0 & 1 & 0 & 0 \\
    & molten iron-d & 0 & 0 & 1 & 1 & 2 & 2 \\
    \bottomrule
  \end{tabular}
\end{table}

\begin{table}[!t]
  \caption{Comparison of RTN and cRTN Scheduling Results in Scenario 2.}
  \label{tab:scenario2}
  \centering
  \begin{tabular}{llcccccc}
    \toprule
    Model & Time slot & 1 & 2 & 3 & 4 & 5 & 6 \\
    \midrule
    \multirow{4}{*}{RTN} 
    & EAF power/MW & 4/3 & 2/3 & 4/3 & 2/3 & 0 & 0 \\
    & molten iron-s & 0 & 1 & 0 & 1 & 0 & 0 \\
    & waiting & 0 & 1 & 0 & 1 & 0 & 0 \\
    & molten iron-d & 0 & 0 & 1 & 1 & 2 & 2 \\
    \midrule
    \multirow{4}{*}{cRTN}
    & EAF power/MW & 4/3 & 4/3 & 4/3 & 0 & 0 & 0 \\
    & molten iron-s & 0 & 1 & 1 & 0 & 0 & 0 \\
    & waiting & 0 & 1 & 1 & 0 & 0 & 0 \\
    & molten iron-d & 0 & 0 & 1 & 2 & 2 & 2 \\
    \bottomrule
  \end{tabular}
\end{table}

\begin{figure}[!t]
  \centering
  \includegraphics[width=3.49in]{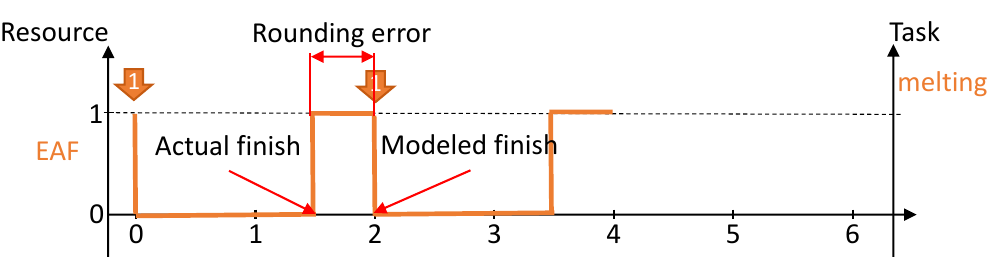}
  \caption{An illustration of the rounding error of the RTN model.}
  \label{fig_rounding_error}
\end{figure}

In Scenario 1, the first four time slots have lower electricity prices, which exactly matches the time needed to produce 2 batches at nominal power. Both RTN and cRTN yielded identical scheduling results, validating that cRTN maintains RTN's expressiveness, particularly in how our designed output and input tasks preserve the discrete nature of batch-based processing even after resource continualization.

In Scenario 2, while the first three time slots have lower electricity prices, RTN still requires four time slots to complete production due to rounding errors (Fig.~\ref{fig_rounding_error}). This occurs because in RTN, a resource (EAF) can only be occupied by one task within a time slot. Therefore, in time slot 2, although the first batch's production could be completed at the midpoint, the second batch's production can only begin in the next time slot. In contrast, cRTN can complete the production of 2 batches within three time slots by operating at 4/3 of nominal power, thereby reducing the electricity cost for the second batch by 25\%.

\section{Case Study}\label{sec_numerical}

In this section, we demonstrate a more realistic application of our cRTN model and validate through numerical results that our method can significantly reduce computation time while maintaining model accuracy. We used Gurobi (V12.0.0)~\cite{gurobi}
and MATLAB (R2024b)~\cite{matlab}
with YALMIP~\cite{Lofberg2004}
to solve the optimization problems. Computation was executed on a workstation with an Intel Core i9-10900X CPU (3.7 GHz) and 128 GB RAM. The default gap setting for GUROBI is 1e-4. Code and data are available at \cite{rick10119_crtn_2024}.

We borrow the scenario of the intraday production schedule of a steel plant under spot electricity prices from \cite{zhang_cost-effective_2017} and use the same parameters for comparison with the legacy RTN model. The steel plant production process to be modeled includes four stages (Fig.~\ref{fig_rtn_process}), namely, the electric arc furnace (EAF), argon oxygen decarburization (AOD), ladle furnace (LF), and continuous casters (CC), the original technical parameters of which are listed in Table~\ref{tab_parameter}. All four stages are typical discrete processes carried out in batches, and each batch, once started, cannot be interrupted. The power of the EAF can be adjusted between 75\% and 125\% of the nominal power, providing additional flexibility. Without loss of generality, we do not consider equipment maintenance time or coupling between parallel production lines, as our model does not make significant improvements in these areas and instead uses existing modeling techniques.

We set the time slot length to 5 minutes, which is consistent with the pricing intervals of some real-time markets. According to the proposed modeling method, the model parameters are listed in Table~\ref{tab_model_parameter} after being converted from the original technical parameters, where the processing rate represents the progress of tasks within each time slot and VT represents the virtual tasks. For example, the processing rate of task waiting (3) is 1/3 in operating state 2, which means that the process needs to be in this state for 3 time slots to complete the transportation between the EAF and AOD (the shortest waiting time is the transportation time, (3-1)*5=10 minutes). The parameters corresponding to the idle state ($k = 0$) are 0.

\begin{table}[!t]
  \caption{Parameters of the original steel plant.}
  \label{tab_parameter}
  \centering
  \begin{tabular}{lllll}
    \toprule
    process                   & EAF & AOD & LF  & CC \\ \midrule
    nominal power (MW)        & 85  & 2   & 2   & 7  \\
    processing time (minute)  & 80  & 75  & 35  & 50 \\
    transfer time (minute)    & 10  & 5   & 10  & /  \\
    max waiting time (minute) & 120 & 120 & 120 & /  \\ \bottomrule
  \end{tabular}
\end{table}

\begin{table}[!t]
  \renewcommand{\arraystretch}{0.8}
  \caption{Parameters of the Continuous RTN Model.}
  \label{tab_model_parameter}
  \centering
  \begin{tabular}{lllll}
    \toprule
  process              & task (index)                    & \begin{tabular}[c]{@{}l@{}}operating\\ state (k)\end{tabular} & \begin{tabular}[c]{@{}l@{}}processing\\ rate (/$\delta$)\end{tabular} & \begin{tabular}[c]{@{}l@{}}operating\\ power (MW)\end{tabular} \\
  \midrule
  \multirow{6}{*}{EAF} & \multirow{3}{*}{processing (1)} & 0                          & 0                          & 0                          \\
                       &                                 & 1                          & $75\% * (5/75)$              & 75\%$*85$                  \\
                       &                                 & 2                          & $125\% * (5/75)$             & 125\%$*85$                 \\ \cline{2-5}
                       & \multirow{3}{*}{waiting (3)}    & 0                          & 0                          & 0                          \\
                       &                                 & 1                          & 1/49                       & 0                          \\
                       &                                 & 2                          & 1/3                        & 0                          \\ \hline
  \multirow{4}{*}{AOD} & \multirow{2}{*}{processing (5)} & 0                          & 0                          & 0                          \\
                       &                                 & 1                          & 1/15                       & 2                          \\ \cline{2-5}
                       & \multirow{3}{*}{waiting (7)}    & 0                          & 0                          & 0                          \\
                       &                                 & 1                          & 1/49                       & 0                          \\
                       &                                 & 2                          & 1/2                        & 0                          \\ \hline
  \multirow{4}{*}{LF}  & \multirow{2}{*}{processing (9)} & 0                          & 0                          & 0                          \\
                       &                                 & 1                          & 1/7                        & 2                          \\ \cline{2-5}
                       & \multirow{3}{*}{waiting (11)}   & 0                          & 0                          & 0                          \\
                       &                                 & 1                          & 1/25                       & 0                          \\
                       &                                 & 2                          & 1/3                        & 0                          \\ \hline
  \multirow{2}{*}{CC}  & \multirow{2}{*}{processing (13)} & 0                          & 0                          & 0                          \\
                       &                                 & 1                          & 1/10                       & 7                          \\ \hline
                       \multirow{2}{*}{VT}                    & \begin{tabular}[c]{@{}l@{}}output \\ (2, 6, 10, 14)\end{tabular}      & 0                          & 0                          & 0                          \\
                       &                                                              & 1                          & 1                          & 0                          \\ \hline
                       \multirow{2}{*}{VT}                    & \begin{tabular}[c]{@{}l@{}}input \\ (4, 8, 12)\end{tabular}      & 0                          & 0                          & 0                          \\
                       &                                                              & 1                          & 1                          & 0                          \\  \bottomrule
                      \end{tabular}
                    \end{table}

\subsection{Task Description and Method Settings}

\begin{figure}[!t]
  \centering
  \includegraphics[width=3.49in]{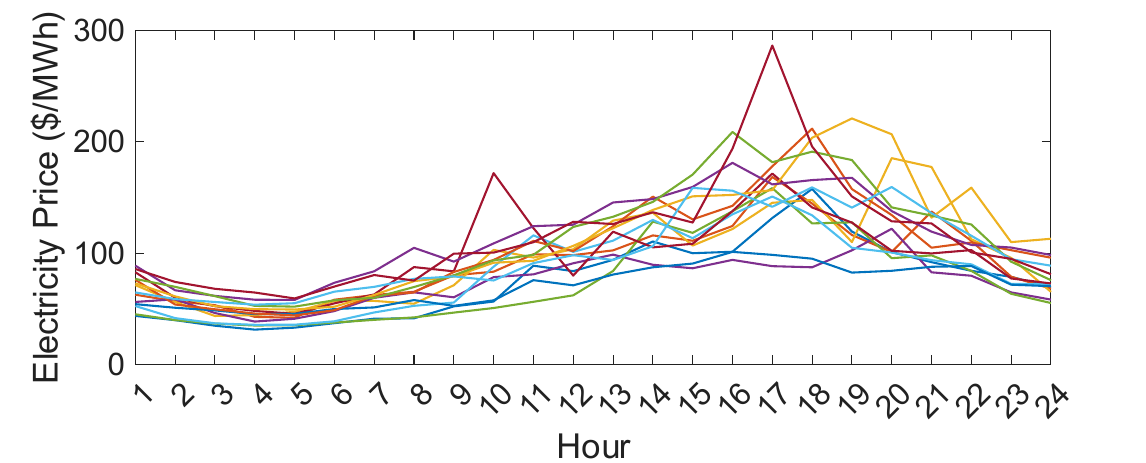}
  \caption{Real-time system electricity prices for PJM on 15 days in July 2022.}
  \label{fig_price}
\end{figure}

In the numerical tests, the model accuracy is measured by the difference in the daily load curve and energy cost given by the proposed method and the optimal scheduling results given by the legacy RTN model, calculated as the RMSE:
$${\rm RMSE}=\sqrt{\frac{\sum^N_{i=1} (y^*_i - y_i)^2}{N}}$$
To consider the differences in electricity prices on different days, we used the real-time system electricity prices of PJM 
on 15 days in July 2022 (Fig.~\ref{fig_price}) and tested the average error. The original hourly prices were linearly interpolated to obtain prices every 5 minutes. We avoided dates with several hours of very close electricity prices because, in these cases, the optimal production scheduling may not be unique. For example, production can be arbitrarily shifted between two hours with the same electricity price without affecting the optimality of the result, which would lead to an overestimation of the model error.

\subsection{Scheduling Results}

\begin{table}[!t]
  \renewcommand{\arraystretch}{1.0}
  \caption{Performance of the cRTN Model. (Heat=8)}
  \label{tab_rmse_models}
  \centering
  \begin{tabular}{llll}
    \toprule
    Model                      &
    \begin{tabular}[c]{@{}c@{}}Rounding \\ error (RMSE)  \end{tabular} &
    \begin{tabular}[c]{@{}c@{}}Average daily \\ energy cost (\$) \end{tabular} &
    \begin{tabular}[c]{@{}c@{}}Mean solution\\ time (minutes) \end{tabular}                                                                      \\ \midrule
    RTN (gap=0.1)                        & -     & 54571                    & -                    \\
    RTN (gap=0.0001)                        & 2.13\%     & 50763(-6.9\%)                    & 156.7                    \\
    cRTN (gap=0.0001)                      & 0 & \textbf{50421 (-7.4\%)} & \textbf{17.1 (-89.0\%)} \\\bottomrule
  \end{tabular}
\end{table}

\paragraph{General Performance}
The general performance of cRTN is described in Table~\ref{tab_rmse_models}. When comparing against the unoptimized case with an optimality gap of 0.1 (which serves as our baseline), both the RTN model and our cRTN model with a tighter gap (0.0001) demonstrate significant improvements in energy cost reduction (-6.9\% and -7.4\% respectively). This improvement primarily stems from their ability to find optimal production schedules. However, a key distinction exists: while both models achieve similar cost reductions, the RTN model inherently suffers from rounding errors (2.13\% RMSE), whereas the cRTN model completely eliminates these errors. 
Moreover, cRTN significantly reduces the solution time for the production scheduling problem compared with the legacy RTN modeling method. In particular, using the cRTN method, the solution time can be reduced to 17.1 minutes (-89.0\%) from that of the legacy RTN model (156.7 minutes).

\begin{figure}[!t]
  \centering
  \includegraphics[width=3.0in]{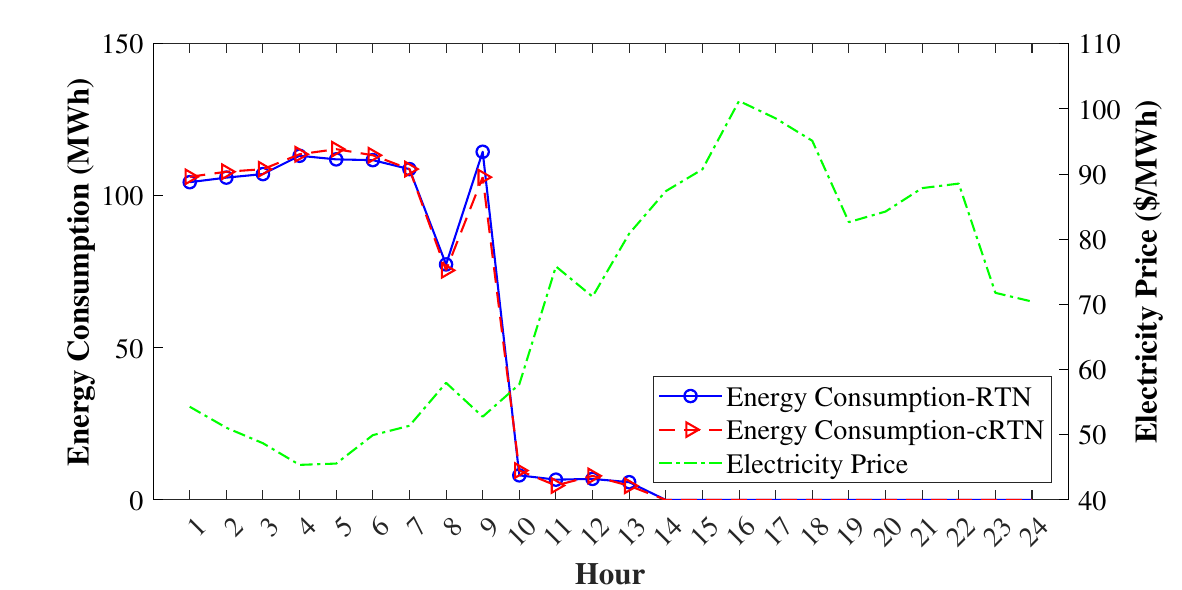}
  \caption{Optimal load curves given by RTN/cRTN for the price of July 15.}
  \label{fig_typical_load}
\end{figure}

\begin{figure}[!t]
  \centering
  \subfloat[]{
    \includegraphics[width=3.0in]{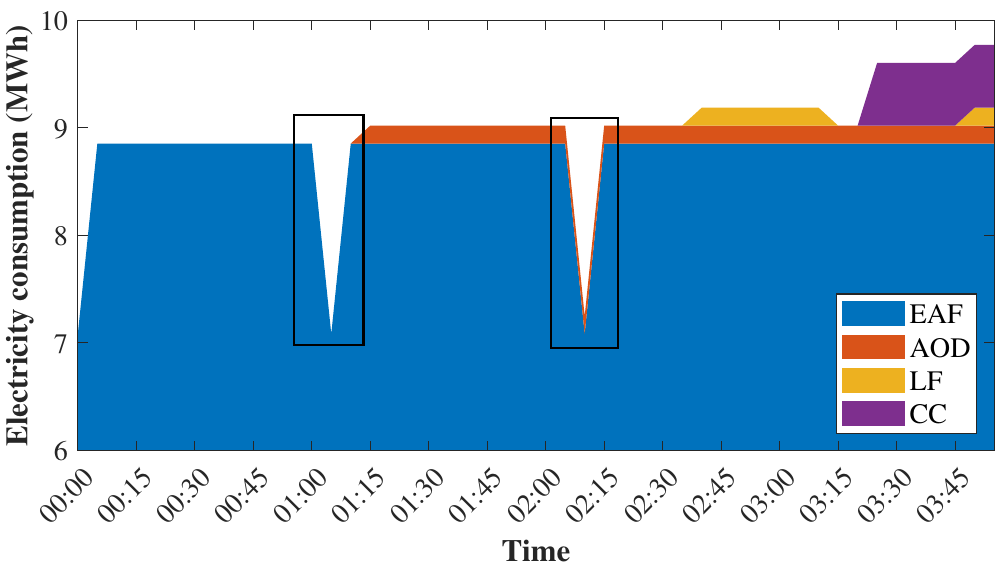}}\\
  \subfloat[]{
    \includegraphics[width=3.0in]{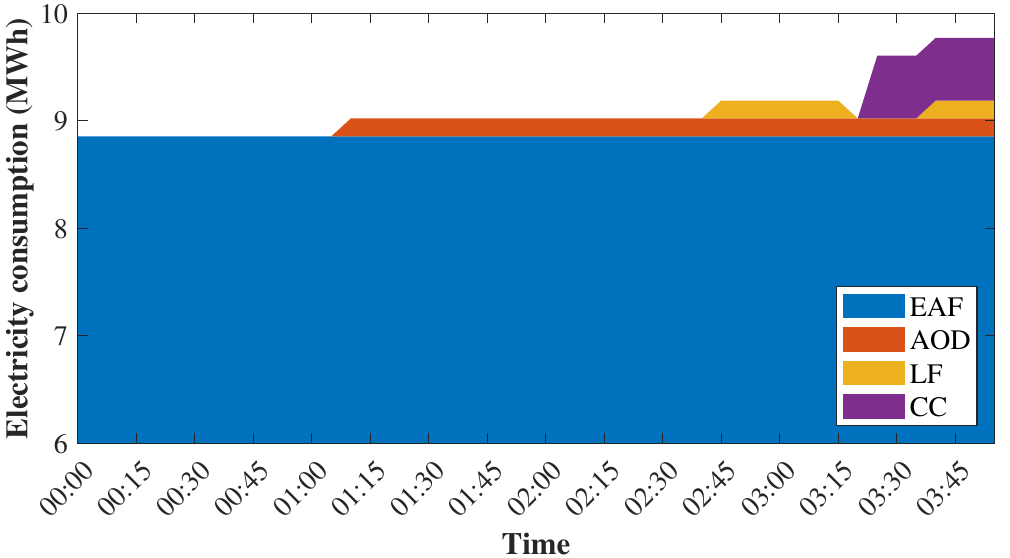}}
  \caption{Breakdown of energy usage every 5 minutes during typical hours (00:00-04:00, July 15): (a) RTN. (b) cRTN. The rounding error (indicated by the black frame) in the RTN model is eliminated in the cRTN, which is the primary source of the difference between the optimal energy usage results given by the cRTN model and those of the RTN model.}
  \label{typical_load_decp}
\end{figure}

\paragraph{Rounding Error} The relative difference between RTN and cRTN's optimal load profiles under 15 different electricity price scenarios is approximately 2.13\%, which can be attributed to the RTN's rounding errors. 
For example, due to the constraints of Equation (\ref{rtn_balance}) under the discrete time horizon framework, the equipment (e.g., EAF) in the RTN model must wait until the next time slot (at 01:05) to start processing the next batch after completing the previous batch (e.g., at 01:01). This means that the device is ``idle'' from 01:01 to 01:05 due to insufficient model representation, which in turn means that the energy use of the RTN model in the corresponding period is not fully optimized. This source of error was reported by Zhang et al.~\cite{zhang_cost-effective_2017} and was referred to as the rounding error. In contrast, the modeling method of the cRTN eliminates the rounding error (see Fig.~\ref{typical_load_decp}), allowing the discrete IP to fully utilize its energy flexibility, resulting in a further reduction of approximately 0.5\% in energy costs. The impact of the rounding error on the energy cost reduction will be more significant, if the length of the time slot is longer. To visualize these differences, we provide the energy use curves of the iron plant under the optimal production schedules given by various methods on a typical day (July 15) in Fig.~\ref{fig_typical_load}.

\begin{figure}[!t]
  \centering
  \includegraphics[width=3.0in]{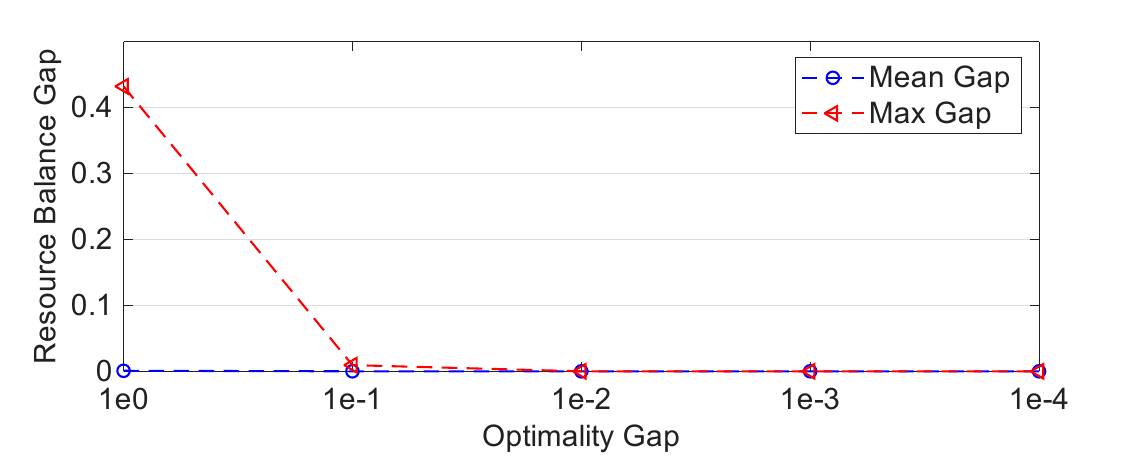}
  \caption{Gap between the two sides of Eq. (12) with different optimality gap settings for the solver. The maximum gap is smaller than the optimality GAP setting for the solver in all of the cases, which means that the two sides of Eq. (12) are equal in our numerical experiments.}
  \label{fig_resource_balance_gap}
\end{figure}

To validate our treatment of the resource balance constraint (Eq. 12), we analyzed the gap between its two sides under different optimality gap settings. As shown in Fig.~\ref{fig_resource_balance_gap}, the maximum gap between the two sides is consistently smaller than the solver's optimality gap setting. For instance, when the optimality gap is set to 1, the maximum resource balance gap is only 0.4. With stricter optimality gaps of 0.1 or smaller, the maximum resource balance gap becomes negligible (less than $10^{-6}$), demonstrating that our relaxation of the equality constraint to an inequality constraint does not compromise the model's accuracy.

\begin{figure}[!t]
  \centering
  \includegraphics[width=3.0in]{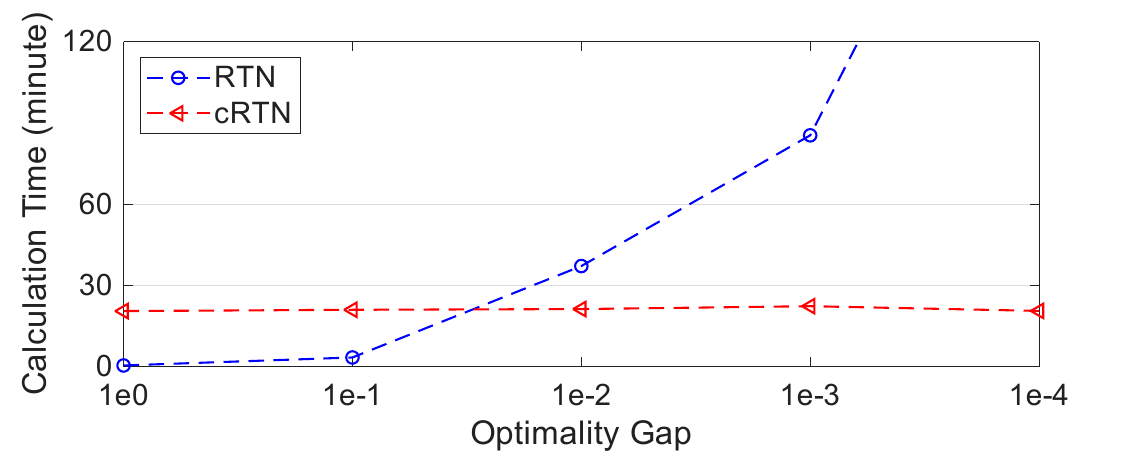}
  \caption{Solution time comparison between RTN and cRTN models under different optimality gap settings.}
  \label{fig_solution_time}
\end{figure}

\paragraph{Impact of Optimality Gap} Fig.~\ref{fig_solution_time} illustrates this performance comparison, showing solution times for both models across different gap settings.
For the RTN model, solution time increases exponentially as the optimality gap requirement becomes stricter. With gaps below 1e-2, computation time exceeds one hour, making it impractical for real-world applications requiring high precision.
The cRTN model demonstrates remarkably stable performance across different gap settings. Even with very strict gap requirements (e.g., 1e-4), the increase in solution time remains marginal, typically staying within practical limits.
From the perspective of solution quality, under a fixed time limit of 30 minutes, cRTN consistently achieves gaps smaller than 1e-4, ensuring high solution quality. In contrast, RTN can only reach gaps around 1e-2 within the same time constraint.
The superior convergence properties of cRTN make it particularly suitable for industrial applications where both computational efficiency and solution precision are critical requirements.

\subsection{Model Scalability}

\begin{figure}[!t]
  \centering
  \includegraphics[width=3.0in]{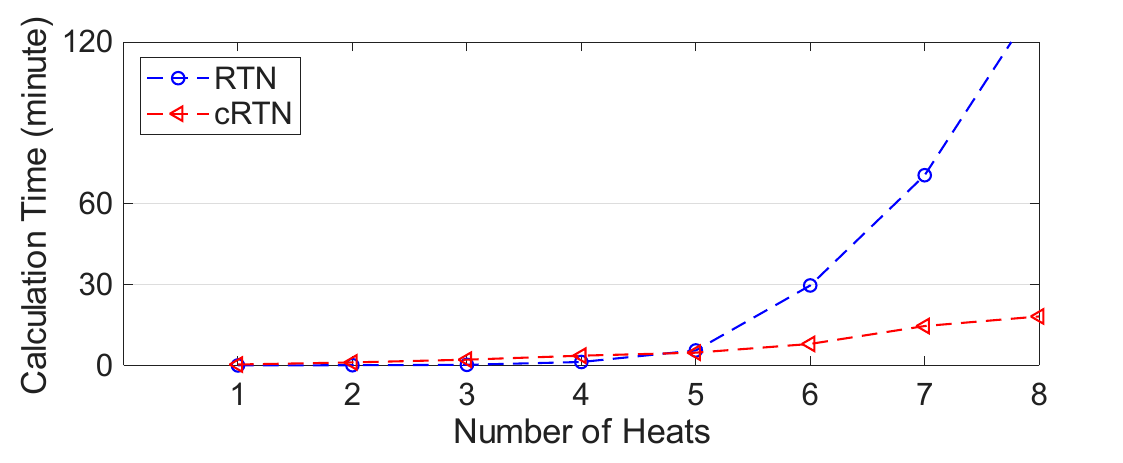}
  \caption{Computation time with respect to the production target (number of heats).}
  \label{fig_calculation_time}
\end{figure}

\begin{table}[!t]
  \caption{Numbers of Variables (Presolved)}
  \label{tab_nofv_models}
  \centering
  \begin{tabular}{lll}
    \toprule
    Model      & \begin{tabular}[c]{@{}l@{}}Binary\\      variables\end{tabular} & \begin{tabular}[c]{@{}l@{}}Continous\\      variables\end{tabular} \\ \midrule
    cRTN       & 4197                       & 8233                       \\
    RTN-5heats & 11513                      & 1262                       \\
    RTN-6heats & 13690                      & 1516                       \\
    RTN-7heats & 15870                      & 1772                       \\
    RTN-8heats & 18046                      & 2025                       \\
    RTN-9heats & 20225                      & 2280                       \\ \bottomrule
  \end{tabular}
\end{table}

Fig.~\ref{fig_calculation_time} shows the solution time (averaged over the 15 price scenarios) of the optimal production scheduling problem based on the cRTN model and the legacy RTN model. When the number of heats reaches 10, the cRTN-based model can still be effectively solved in an hour, while the solution time of the RTN-based model exceeds 2 hours when the number of heats is more than 7, which is not practical for demand response applications. The solution time of the RTN shows close to an exponential increase in solution time. The reason is, as mentioned before, under the legacy RTN modeling method, the scale of the model grows linearly with increasing production objectives (the number of heats). When heat $=9$, binary variables increase to over 20000 (Table~\ref{tab_nofv_models}, presolved using GUROBI). For MILP problems, this usually means exponential growth in the solution time. The number of variables, 4197, in the proposed cRTN is far less than that in the legacy RTN method. When heat $=9$, the proposed cRTN model has 4197 binary variables, the same as when heat $=1$, and this number does not increase with the production target. Nevertheless, the solution time of the cRTN does increase with the addition of heats (Fig.~\ref{fig_calculation_time}) because the solution time is also related to the parameter settings. However, the increase in solution time for the cRTN model is much slower than that for the RTN model.

\section{Conclusions and Discussion}\label{sec_conclusion}

This paper proposes the cRTN model for computationally efficient modeling of discrete IPs. cRTN systematically reconstructs the mathematical formulation of discrete IP models through 1) decoupling the modeling from batches by using continuous variables to represent resource quantities and task operations, and 2) introducing carefully designed binary variables to model process states, thus preserving the discrete nature of IPs. The numerical results demonstrate two significant improvements:

1) The cRTN model maintains high accuracy compared to the legacy RTN model while achieving a dramatic reduction in solution time ($>89.0\%$). 

2) By resolving the rounding error issue inherent in legacy RTN models, cRTN can achieve up to 25\% reduction in energy costs under certain scenarios.

This work should be considered a first effort to reformulate the RTN model since its inception. Our novel formulation replaces the binary variables for task execution in the legacy RTN model with continuous variables, an approach initially explored in our continuous IP modeling study~\cite{lyu_lstn_2023}. Most importantly, this reformulation establishes a unified framework that bridges discrete IPs and continuous IPs. By constructing a continuous RTN model based on the linearized version of the STN model, we demonstrate that continuous production processes are essentially a special case of discrete ones. This theoretical unification reveals that discrete IP models can be systematically derived by adding appropriate constraints to continuous IP models, providing a new understanding of IP modeling.

Computational efficiency is crucial for real-world demand response implementation, particularly in scenarios requiring frequent rescheduling or coordination among multiple industrial facilities. While this paper focuses on scheduling the production of an individual industrial facility, the aggregation of multiple IPs' flexibility using methods such as clustering could be valuable for power system planning and operation~\cite{6684593}. Notably, the resource conversion and task execution processes of the proposed cRTN model, after being made continuous, are mathematically more conducive to potential applications of the model, such as flexibility aggregation~\cite{lyu2024approximating} and real-time optimization~\cite{chen_real-time_2024}, which are also directions for future research.




\appendices
\section{Resource Balance in STN and cRTN Models}\label{app_stn_crtn_comparison}

In the STN model~\cite{lyu_lstn_2023}, the resource balance is formulated to track the evolution of material states over time. Let $S_{ti}$ (kg) denote the amount of material $i$ at the end of time interval $t$. The model uses $\Delta t_{tik}$ (h) to represent the continuous time duration that task $i$ operates at point $k$ within time period $t$. For each task $i$, let $G_{ik}$ and $C_{ik}$ (kg/h) denote the material production and consumption rates at operating point $k$, respectively. The evolution of material states is governed by three distinct balance equations, corresponding to feedstock, intermediate products, and final products:

For feedstock:
\begin{equation}\label{primal_constraint_changeofS1}
  S_{ti} = S_{(t - 1)i} - \sum_{k \in K_{i+1}} C_{(i+1)k} \Delta t_{t(i+1)k}
\end{equation}

For intermediate products:
\begin{equation}\label{primal_constraint_changeofS2}
    S_{ti} = S_{(t - 1)i} + \sum_{k \in K_{i}} G_{ik} \Delta t_{tik} - \sum_{k \in K_{i+1}} C_{(i+1)k} \Delta t_{t(i+1)k}
\end{equation}

For final products:
\begin{equation}\label{primal_constraint_changeofS3}
  S_{ti} = S_{(t - 1)i} + \sum_{k \in K_{i}} G_{ik} \Delta t_{tik}
\end{equation}

These equations can be generalized using a more compact notation. By introducing $g_{r, i, k}$ to represent the resource consumption/production rate of task $i$ on resource $r$ at operating point $k$ (where $g_{r, i, k} > 0$ indicates production and $g_{r, i, k} < 0$ indicates consumption), the resource balance can be expressed in a unified form:

\begin{equation}\label{stn_balance}
  R_{r, t} = R_{r, t-1} + \sum_{i} \sum_{k} g_{r, i, k} \Delta t_{tik} \quad \forall r, t
\end{equation}

where $R_{r, t}$ represents the amount of resource $r$ at time $t$. This provides a unified formulation for modeling resource balance in the STN model and the cRTN model, while for the cRTN model, more constraints are added to describe the discrete nature of the production process. This indicates that the cRTN model is a more general form of the STN model, and that continuous IPs are a degenerated case of discrete IPs.


\section*{Acknowledgment}
Generative artificial intelligence was used to enhance grammar and readability.


\ifCLASSOPTIONcaptionsoff
  \newpage
\fi



\bibliographystyle{IEEEtran}
\bibliography{reference}

\begin{IEEEbiography}[{\includegraphics[width=1in,height=1.5in,clip,keepaspectratio]{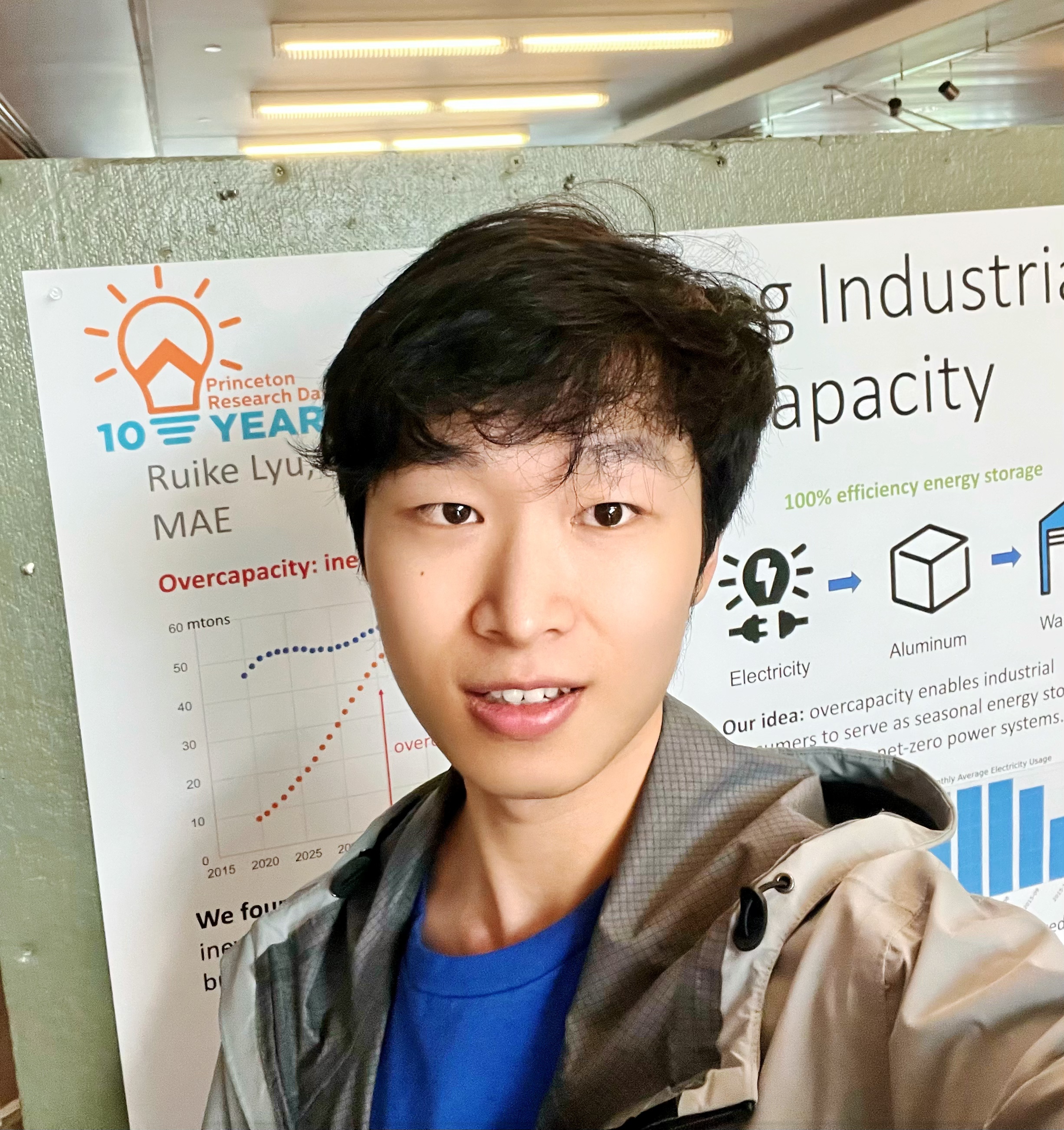}}]{Ruike Lyu}

  received the bachelor's degree in electrical engineering from Tsinghua University, Beijing, China, in 2021, where he is currently pursuing the Ph.D. degree, advised by Prof. Chongqing Kang and Prof. Hongye Guo. Since February 2025, he has been a visiting scholar at Princeton University, advised by Prof. Jesse Jenkins. His research focuses on demand-side flexibility from electric vehicles and industrial loads, particularly their integration into power markets. Ruike has received several awards for his work, including Best Paper/Presentation at CEEPE 2024, EECT 2025, and PSSGT 2025. He was also awarded Best Presentation at the IEEE PES Ph.D. Dissertation Challenge in 2025.
  
\end{IEEEbiography}

\begin{IEEEbiography}[{\includegraphics[width=1in,height=1.25in,clip,keepaspectratio]{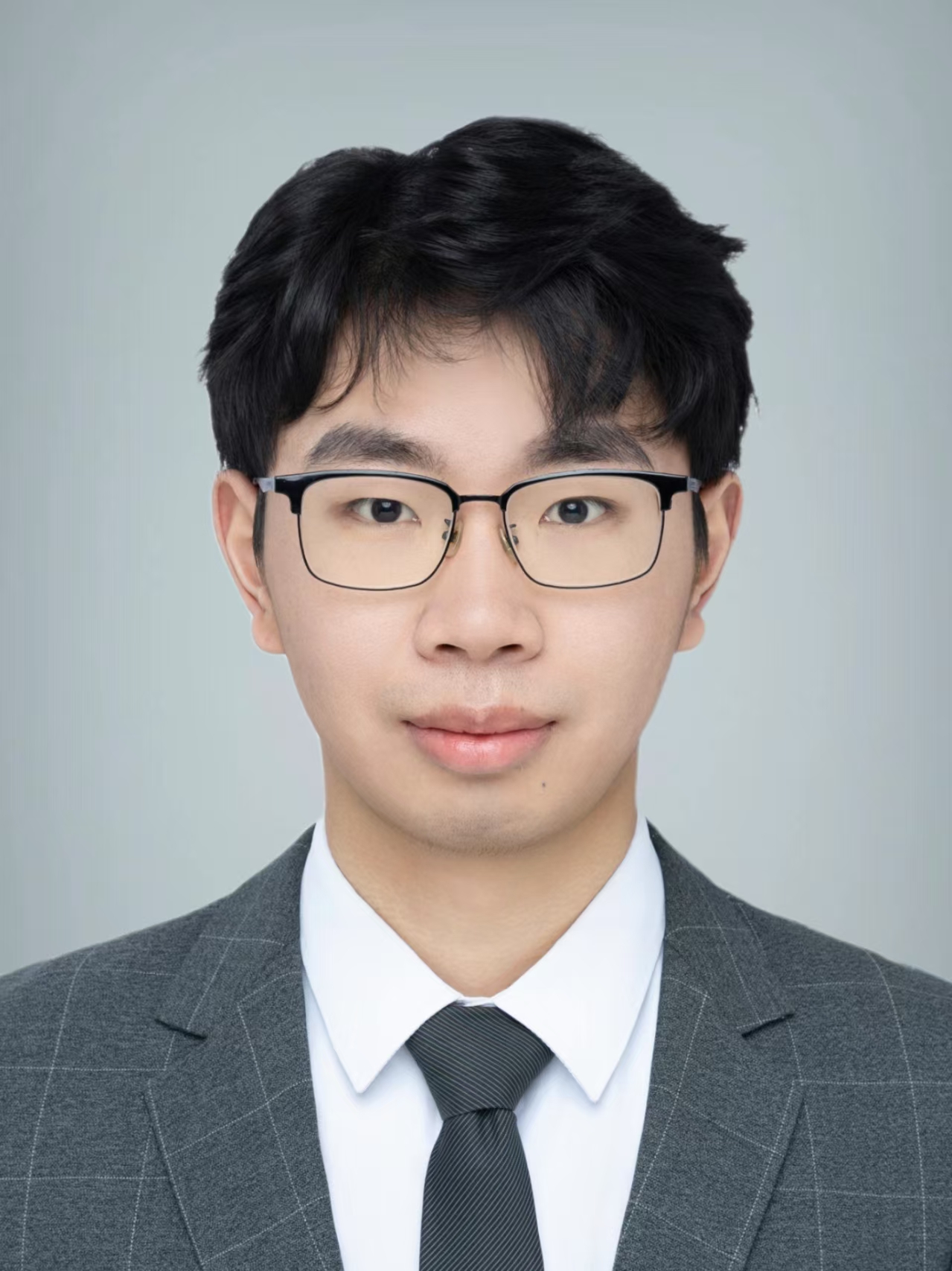}}]{Xiangbo Su}

 received the bachelor's degree in electrical engineering from Huazhong University of Science and Technology, Wuhan, China, in 2022. He is currently pursuing the master's degree in the Department of Electrical Engineering at Tsinghua University, Beijing, China. His research focuses on the potential assessment of industrial consumers.
  
\end{IEEEbiography}

\begin{IEEEbiography}[{\includegraphics[width=1in,height=1.25in,clip,keepaspectratio]{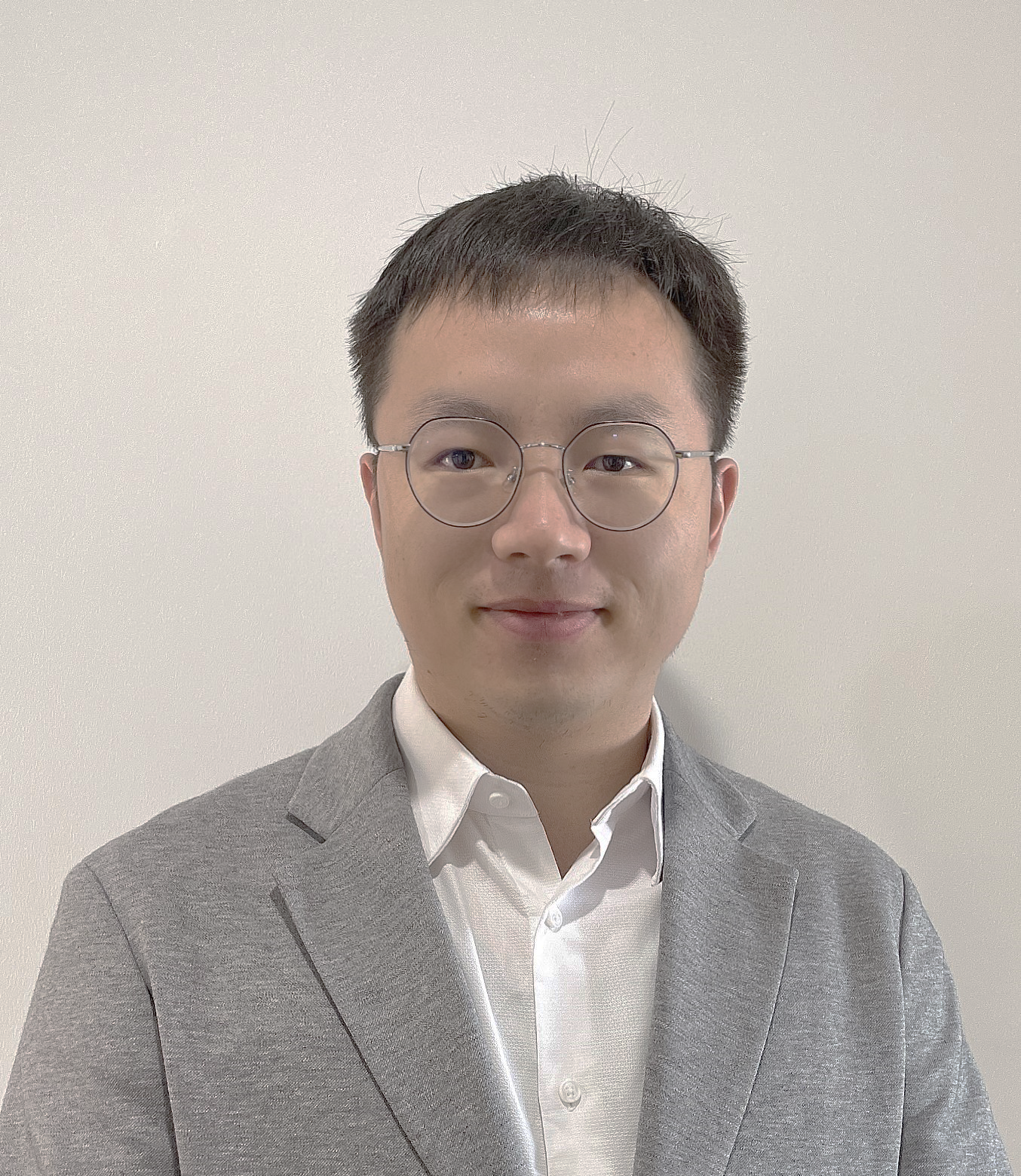}}]{Ershun Du}

 (Member, IEEE) received the B.S. and
Ph.D. degrees from the Department of Electrical
Engineering, Tsinghua University in 2013 and 2018,
respectively. He is currently an Associate Professor
with Tsinghua University. His research interests
include low-carbon energy policy and climate change,
renewable energy uncertainty analysis, power system
economics and planning.
  
\end{IEEEbiography}

\begin{IEEEbiography}[{\includegraphics[width=1in,height=1.25in,clip,keepaspectratio]{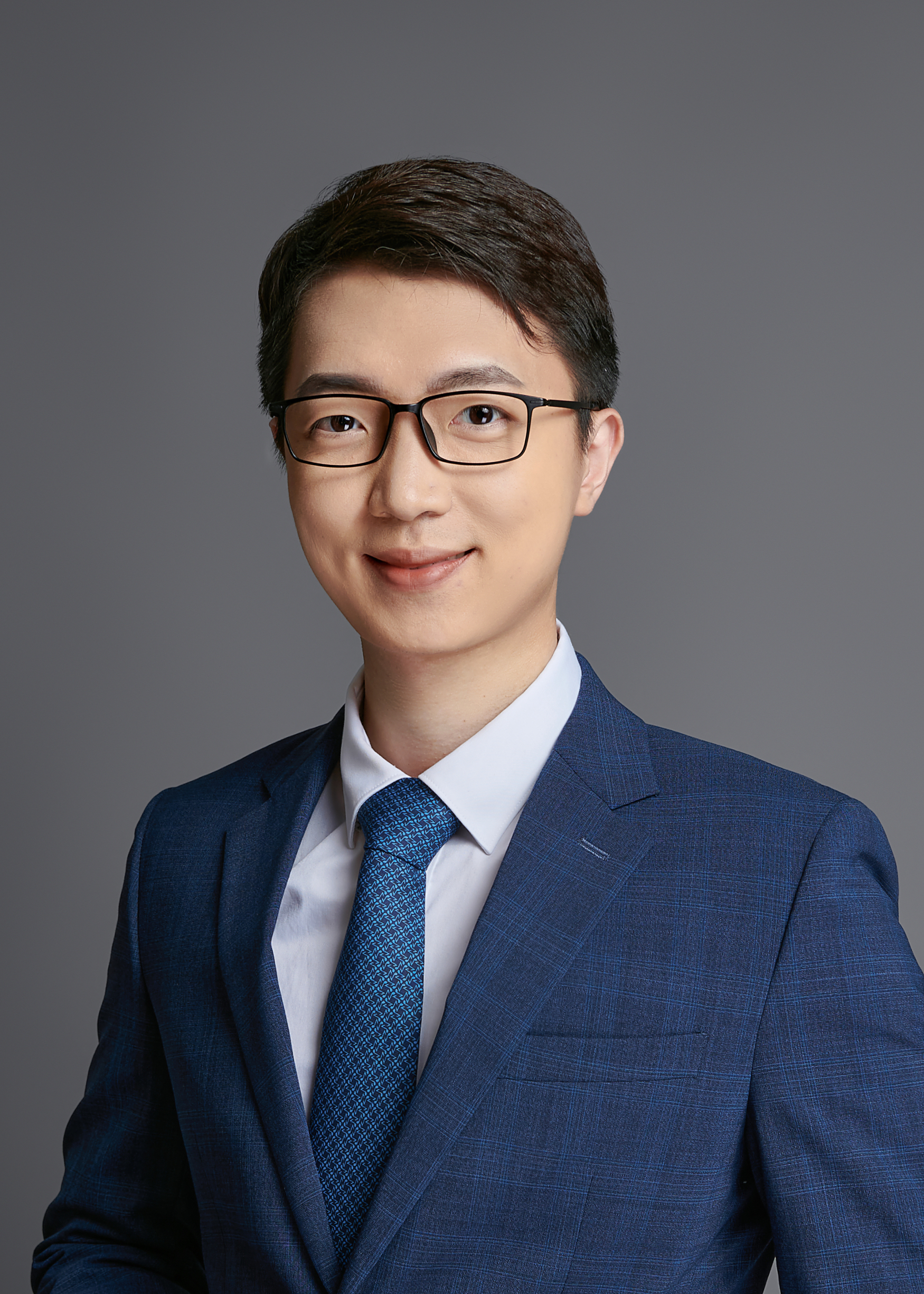}}]{Hongye Guo}

  (S'15-M'20) received the B.S. and Ph.D. degrees in electrical engineering from Tsinghua University, Beijing, China, in 2015 and 2020, respectively. He was a post-doctor in electrical engineering from Tsinghua University from 2020 to 2022. He was a visiting student researcher with Stanford University, CA, USA, in 2018, and with Illinois Institute of Technology, Chicago, IL, USA, in 2019. He is currently an associate professor at Tsinghua University. His research interests include electricity markets, demand-side flexibility, and machine learning applications in power markets.
  
\end{IEEEbiography}

\begin{IEEEbiography}[{\includegraphics[width=1.0in,height=1.25in,clip,keepaspectratio]{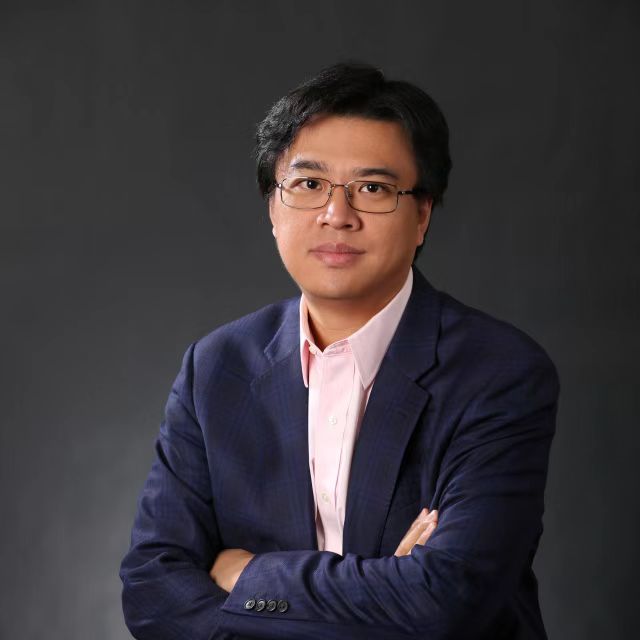}}]{Qixin Chen}
(Senior Member, IEEE) received the Ph.D. degree from the Department of Electrical Engineering, Tsinghua University, Beijing, China, in 2010. His research interests include electricity markets, power system economics and optimization, low-carbon electricity, and power generation expansion planning.
\end{IEEEbiography}

\begin{IEEEbiography}[{\includegraphics[width=1in,height=1.25in,clip,keepaspectratio]{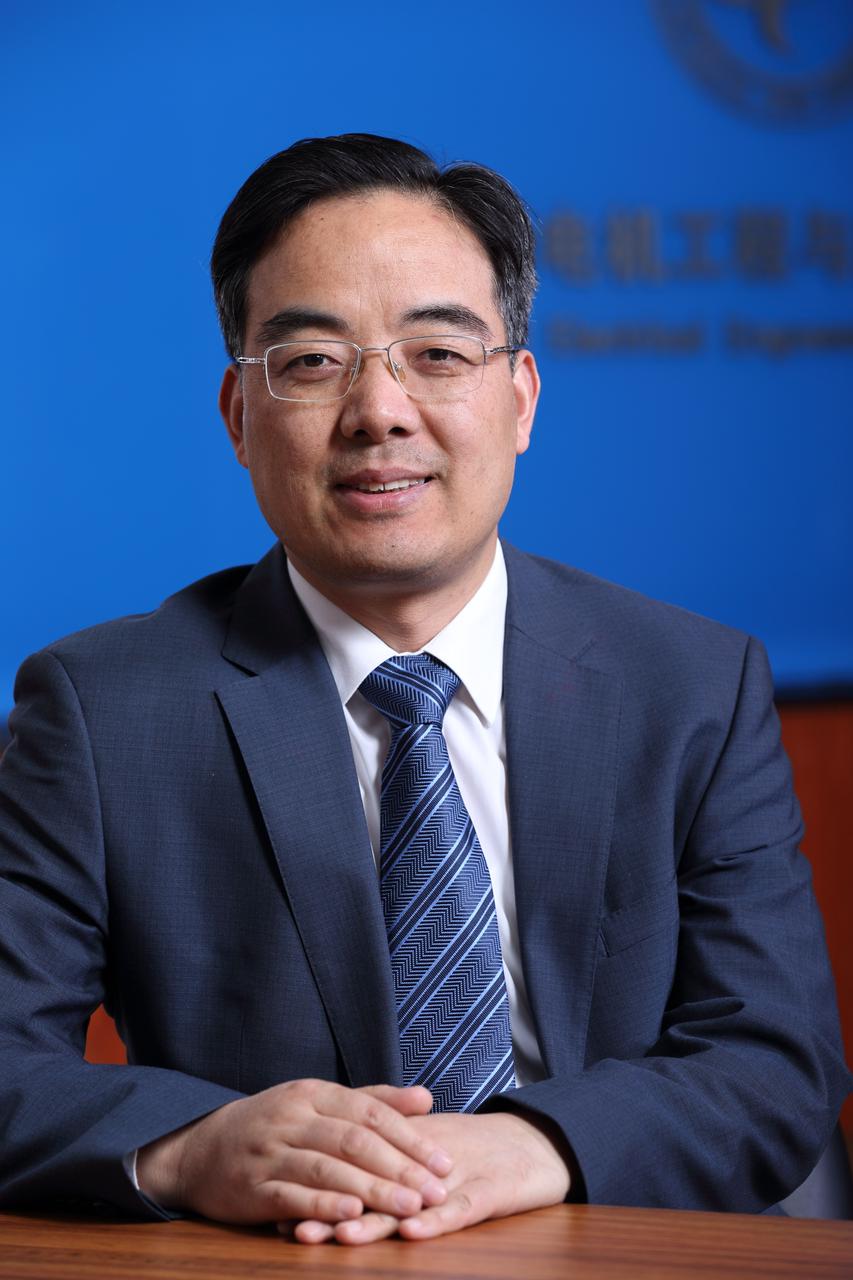}}]{Chongqing Kang}

 (Fellow, IEEE) received the
 Ph.D. degree from the Department of Electrical
 Engineering, Tsinghua University, Beijing, China,
 in 1997, where he is currently a Professor. His
 research interests include power system planning,
 power system operation, renewable energy, low-carbon electricity technology, and load forecasting.
  
\end{IEEEbiography}

\end{document}